\documentclass[aps,twocolumn,floats,showpacs,amsmath,superscriptaddress,prl]{revtex4-1}
\usepackage{graphicx}
\usepackage{dcolumn}
\usepackage{bm}
\usepackage{amsmath}
\usepackage{amsfonts}
\usepackage{amssymb}
\usepackage{xcolor}
\usepackage{soul}
\usepackage{epstopdf}
\usepackage{epsfig}
\usepackage{lipsum}

\newcommand{\de}[1]{\textcolor{black}{#1}}
\newcommand{\jb}[1]{\textcolor{black}{#1}}
\newcommand{\co}[1]{\textcolor{black}{#1}}

\begin{document}



\title{The fate of interaction-driven topological insulators under disorder}

\date{\today}

\author{Jing Wang}
\affiliation{Department of Modern Physics, University of Science and
Technology of China, Hefei, Anhui 230026, P.R. China}
\affiliation{Institute for Theoretical Solid State Physics,
IFW Dresden, Helmholtzstr. 20, 01069 Dresden, Germany}
\affiliation{Department of Physics, Tianjin University, Tianjian 300072, P.R. China}

\author{Carmine Ortix}
\affiliation{Institute for Theoretical Solid State Physics,
IFW Dresden, Helmholtzstr. 20, 01069 Dresden, Germany}
\affiliation{Institute for Theoretical Physics, Center for Extreme Matter
and Emergent Phenomena, Utrecht University, Princetonplein 5, 3584 CC Utrecht, Netherlands}
\author{Jeroen van den Brink}
\affiliation{Institute for Theoretical Solid State Physics,
IFW Dresden, Helmholtzstr. 20, 01069 Dresden, Germany}
\affiliation{Institute for Theoretical Physics, TU Dresden,
01069 Dresden, Germany}
\author{Dmitry V. Efremov}
\affiliation{Institute for Theoretical Solid State Physics,
IFW Dresden, Helmholtzstr. 20, 01069 Dresden, Germany}

\begin{abstract}
We analyze the effect of disorder on the weak-coupling instabilities of quadratic band crossing point (QBCP) in
two-dimensional Fermi systems, which,  in the clean limit, display interaction-driven topological insulating
phases. In the \jb{framework of a} renormalization group procedure, which treats fermionic interactions
and disorder on the same footing, we test all possible instabilities and identify the corresponding ordered
phases in the presence of disorder for both single-valley and two-valley QBCP systems.
We find that disorder generally \co{suppresses the critical temperature at which the interaction-driven topologically non-trivial order sets in.
Strong disorder can also cause a topological phase transition into a topologically trivial insulating state.
}
\end{abstract}

\maketitle

The study of topological phases of matter is one of the most active
research areas in  contemporary condensed matter physics. The explanation of the quantum Hall
effect in terms of the topological properties of the Landau levels \cite{Laughlin1981PRB,Thouless1982PRL} in the 1980's
triggered an intense research effort in the theoretical prediction  \cite{Haldane1988PRL,Kane2005PRL,Bernevig2006Science}
and the experimental discovery \cite{Xue2013Science,Konig2007Science} of a plethora of different topologically non-trivial
quantum phases. In two-dimensional (2D) insulating systems only two distinct topological non-trivial phases can be realized according to the
well-established classification of topological insulators and superconductors \cite{Altland1997PRB,Schnyder2008}
: (i) the quantum anomalous Hall state (QAH) \cite{Haldane1988PRL} with a time-reversal symmetry-broken ground state and topologically protected chiral edge states and (ii) the time-reversal
invariant quantum spin Hall (QSH) state \cite{Kane2005PRL,Bernevig2006Science}, which possesses helical edge states with
counter-propagating electrons of opposite spins.

In recent years, attention has gradually shifted from non-interacting topological states of matter
towards {\it interaction-driven topological phases}: many-particle quantum ground-states in which
chiral orbital currents or spin-orbit couplings are spontaneously generated by electronic correlations.
These states of matter possess both conventional order, characterized by an order parameter and a broken symmetry, and protected edge states associated with a topological quantum number.
Interaction-driven QAH and QSH phases were first conceived in the
context of 2D honeycomb lattice Dirac fermions \cite{Raghu2008PRL} assuming sufficiently strong electronic
repulsions although more recent analytical and numerical works question the proposal for
this particular model \cite{Martinez2013PRB,Daghofer2014PRB,Motruk2015PRB,Capponi2015PRB}.

\begin{figure}
\centering
\includegraphics[width=2.6in]{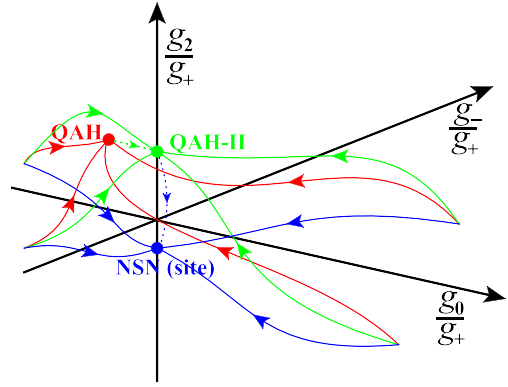}
\vspace{-0.30cm}
\caption{(Color online) Schematic renormalization group flows of the coupling constant ratios
$(g^*_0,g^*_-,g^*_2)/g^*_+$  for the checkerboard lattice in the clean limit (red) and in the
presence of random chemical potential(green and blue). \textcolor{black}{In clean limit the QAH fixed point $(g^*_0,g^*_-,g^*_2)/g^*_+
=(0,-3.73,7.46)$ (red point) corresponds to a QAH state. }In the presence of the weak random chemical potential the coupling
constants flow to \textcolor{black}{the QAH-II fixed point (green) $(g^*_0,g^*_-,g^*_2)/g^*_+ =(0,-0.2,6.5)$ corresponding
again to the QAH state and in the presence of the strong random chemical potential to the NSN fixed point (green) }
$(g^*_0,g^*_-,g^*_2)/g^*_+=(0,0,-1.09)$ corresponding to the NSN (site) state. The dashed lines represent all virtual
trajectories to change the fixed points \textcolor{black}{with impurities}. \textcolor{black}{The relationship between fixed points
QAH and QAH-II is provided in the inset figure of Fig.~\ref{Fig_Phase_diagram_chemical}(a).}}\label{Fig_shcematic_FP_flows_M0}
\end{figure}

\co{On the contrary,} it has been proposed that 2D systems with a quadratic band crossing point (QBCP) are unstable
to electronic correlation because of the finite density of states at the Fermi level  leading to
the possibility of weak-coupling interaction-driven topological insulating phases \cite{Fradkin2009PRL,Venderbos2016,Wu2016}.
And, indeed, QAH and QSH phases generated by electronic repulsions occur both in the checkerboard
lattice \co{model} \cite{Fradkin2009PRL,Vafek2014PRB}, and in
\co{two-valley QBCP models for bilayer graphene}
\cite{Vafek2010PRB,Vafek2010PRB_2}.
A question that naturally arises is whether and how these weak-coupling interaction-driven
states are affected by the presence of disorder, which is well-known to induce prominent phenomena
such as Anderson localization, metal-insulator transition and phase transitions between superconducting
phases \cite{Lee1985RMP,Mirlin2008RMP,Nersesyan1995NPB,Fiete2016PRB,
Efremov11,Efremov13,Korshunov2014}.
\jb{For 2D topological states of matter this question is of particular importance, as the global topological nature of the ground state \co{should} render such states in principle robust against the local effects of disorder \cite{Halperin1982PRB}.}

In this \co{work}, we analyze the fate of the interaction-driven topological insulators in Fermi systems
with a QBCP under the effect of three different  types of disorders, \de{which preserve time reversal symmetry}  \cite{Nersesyan1995NPB}.
Depending on their couplings with fermions we refer to these as random chemical potential, random mass,
and random gauge potential \cite{Stauber2005PRB}. These different sorts of disorders have been shown to give rise to
distinct behaviors of fermionic systems \cite{Fradkin2010ARCM,Altland2002PR,Lee2006RMP, Lee2005Nature,Aleiner2006PRL,
Neto2009RMP,Sarma2011RMP,Kotov2012RMP,Ludwig1994PRB,Furneaux1995PRB,Ye1998PRL,Hasan2010RMP,
Sachdev1999Book,Wang2011PRB}.
\de{In general, the effect of disorder is essential in 2D for itinerant systems since it may lead to localization.}
\jb{Therefore interactions and disorder must be treated on equal footing and we subsequently go beyond a mean-field analysis of}
disorder and employ the perturbative renormalization group (RG) technique \cite{Wilson1975RMP,Polchinski9210046,Shankar1994RMP}.
The renormalization flow procedure starts at high energy, when the ground state is known, and ends with a leading instability, which is characterized by a corresponding fixed point (FP). \textcolor{black}{The analysis of the interplay of the
phases well below $T_c$ is out of  scope of the present paper.}
The central result of our calculations is schematically illustrated in
Fig.~\ref{Fig_shcematic_FP_flows_M0} for the random chemical potential.
With disorder the fixed points evolve to new positions, which correspond to
topological
phase transitions
\co{to trivial insulating states in the strong disorder regime. Moreover, }
\de{the analysis of the evolution of the FP}
shows that disorder
\co{generally suppresses the critical temperature at which the interaction-induced topological insulating states set in.}

\begin{figure}
\hspace{3.50cm}\includegraphics[width=1.65in]{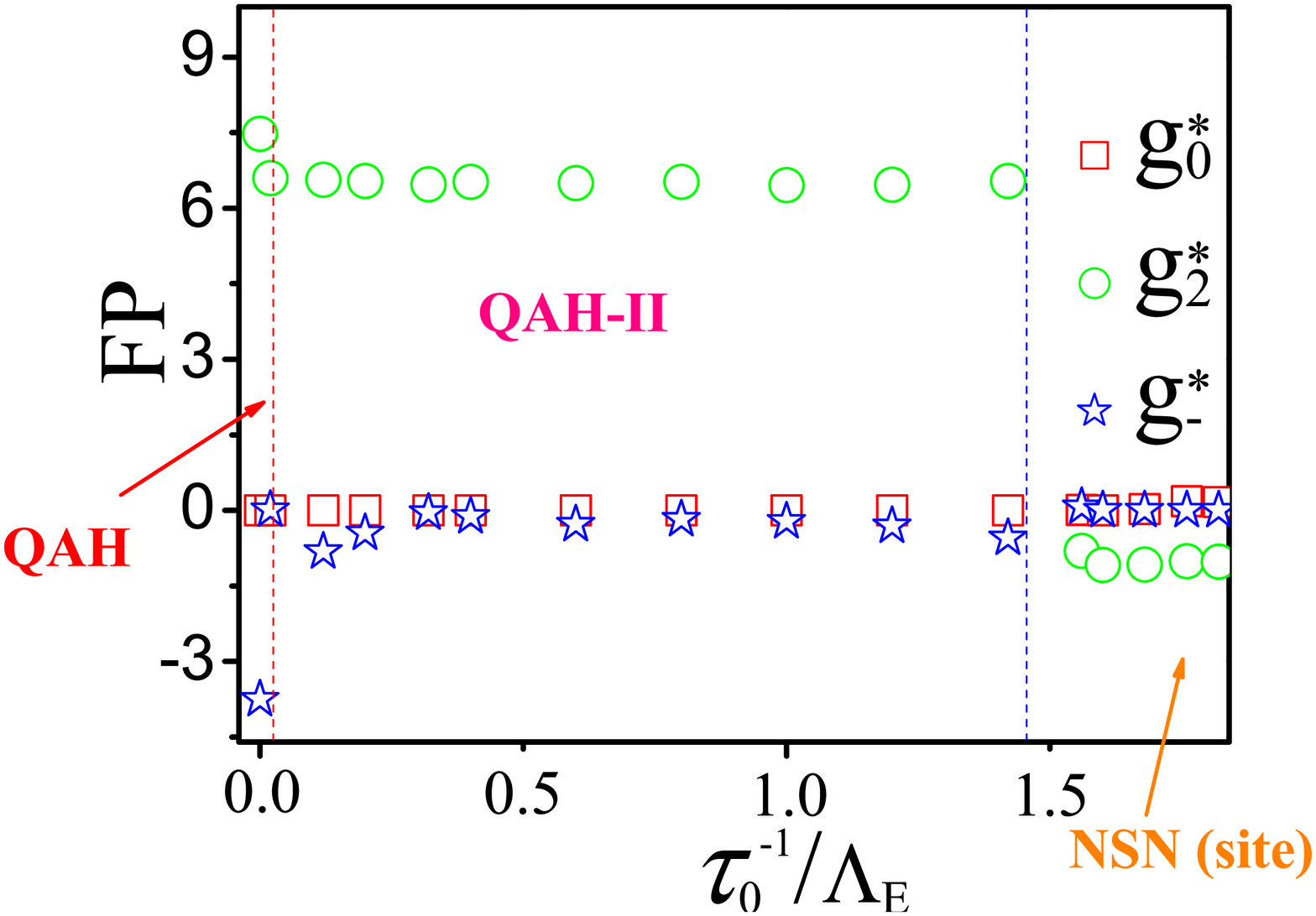}\vspace{-3.1cm}
\includegraphics[width=2.8in]{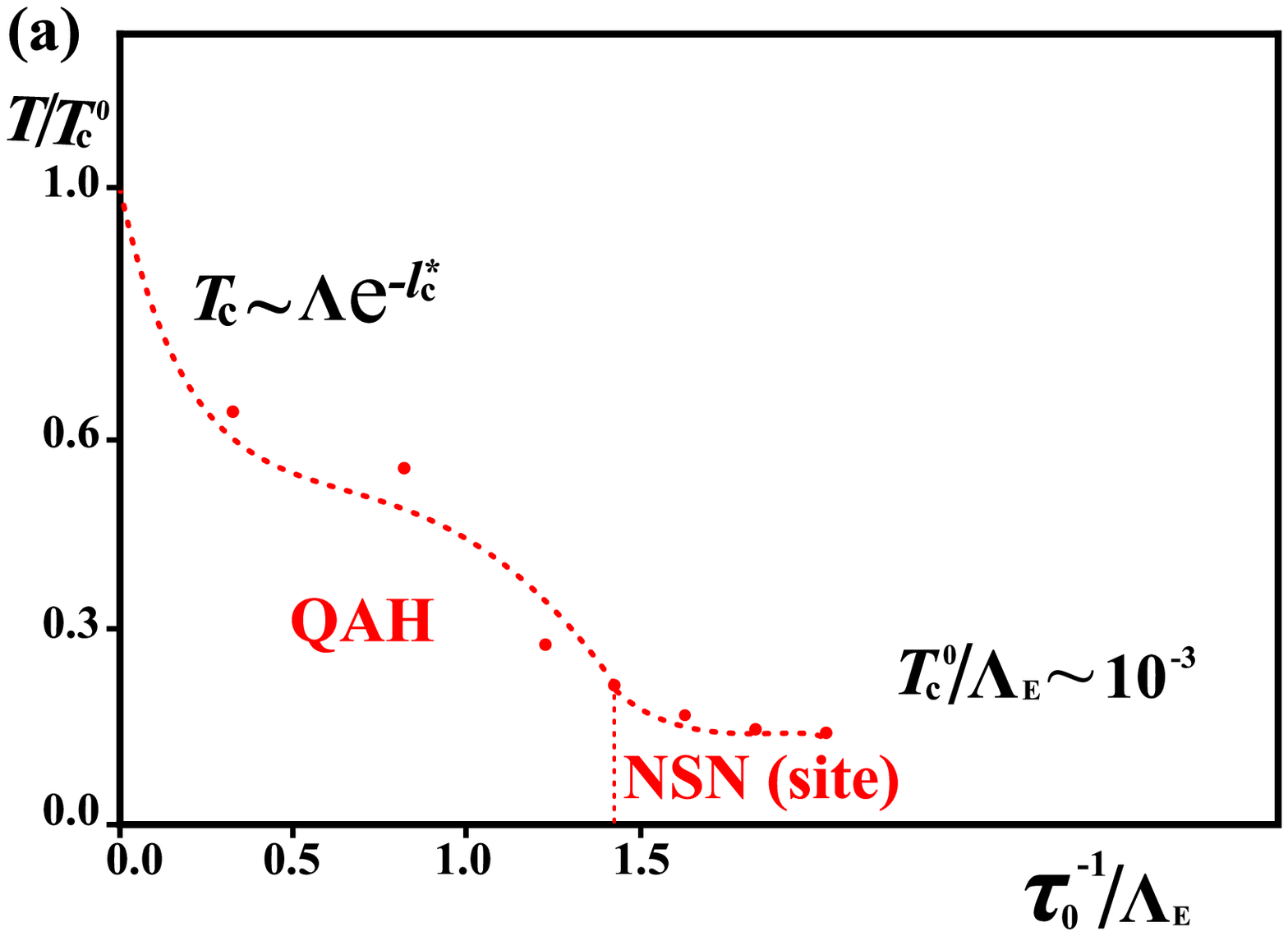}\\
\hspace{3.50cm}\includegraphics[width=1.65in]{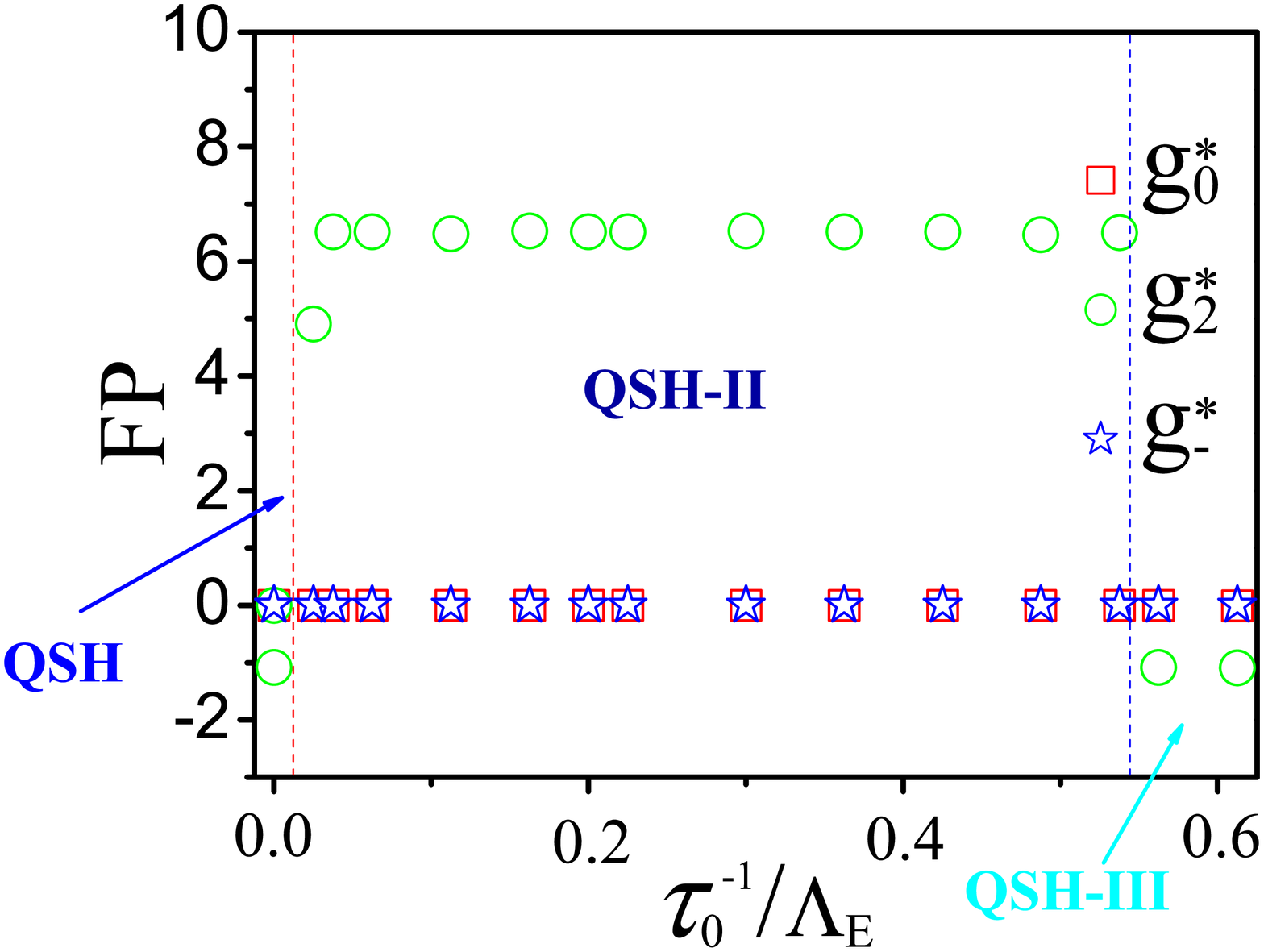}\vspace{-3.1cm}
\includegraphics[width=2.8in]{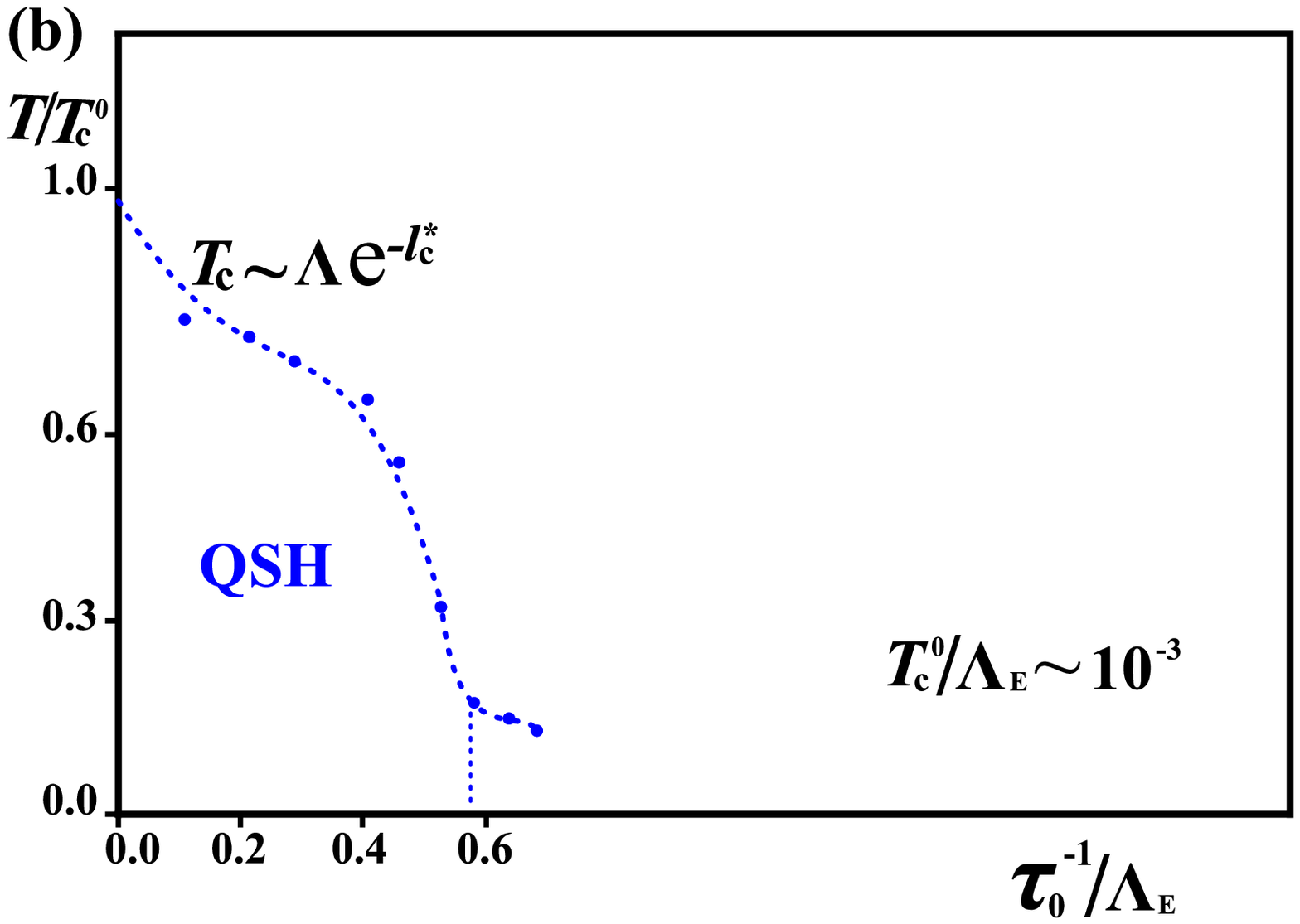}
\vspace{-0.2cm}
\caption{(Color online) Schematic phase diagram for the checkerboard lattice in the presence
of random chemical potential and which in the clean limit give (a): $(g^*_0,g^*_-,g^*_2)=(0,-3.73,7.46)g^+$ and
(b): $(g^*_0,g^*_-,g^*_2)=(0,0,-1.09)g^+$. \textcolor{black}{The other two cases are presented in the SM \cite{Supple_materials}.
Insets: Evolution of the (a) QAH fixed points $(g^*_0,g^*_-,g^*_2)=(0,-3.73,7.46)g_+$ and (b) QSH $(g^*_0,g^*_-,g^*_2)=(0,0,-1.09)$
with increase of the strength of the bare random mass disorder potential for the checkerboard lattice.
The designations QAH and QAH-II (or QSH, QSH-II and QSH-III) FPs show the regimes, where the FPS are
stable with increase of impurity scattering rates. The inset QAH and QSH represent the clean limit case.}}\label{Fig_Phase_diagram_chemical}
\end{figure}

\emph{Checkerboard lattice --}
The low-energy theory of spin one-half fermions
on a checkerboard lattice in the presence of disorder is
described by the Hamiltonian $H = H_0+H_{\mathrm{int}} + H_{\mathrm{dis}}$, where $H_0$ is the
kinetic energy, which is invariant under the $C_{4v}$ point group
and time-reversal symmetry \cite{Fradkin2009PRL}. It reads:
\begin{eqnarray}
H_0&=&\sum_{|\mathbf{k}|<\Lambda}\sum_{\sigma=\uparrow\downarrow}\psi^\dagger_{\mathbf{k}\sigma}
\mathcal{H}_0(\mathbf{k})\psi_{\mathbf{k}\sigma}, \\ \mathcal{H}_0(\mathbf{k})
&=&t_I\mathbf{k}^2 \tau_0+2t_xk_xk_y\tau_1+t_z(k^2_x-k^2_y)\tau_3,
\end{eqnarray}
where $\Lambda$ is the momentum cut-off, while $\psi_{\mathbf{k}\sigma}$ has two components corresponding
to the two sublattices of the checkerboard lattice and  $\tau_i$ are Pauli matrices.
Without loss of generality, we will consider in the remainder the parameter set $t_I =0$ and $t_x=t_z=t$,
which corresponds to a particle-hole symmetric QBCP \cite{Fradkin2009PRL,Vafek2012PRB,Vafek2014PRB}
\textcolor{black}{and the parameter $t$ is rescaled by $1/2m$ (here and below we assume $\hbar=1$).} The interacting part of the
Hamiltonian $H_{\mathrm{int}}$ has the general form \cite{Fradkin2009PRL,Vafek2012PRB,Vafek2014PRB,Wen2008PRB,Fradkin2008PRB}:
\begin{eqnarray}
H_{\mathrm{int}}=\frac{2\pi}{m}\sum_{i =0}^{3}g_i\int d^2\mathbf{x}
\left(\sum_{\sigma=\uparrow\downarrow}\psi^\dagger_\sigma(\mathbf{x})
\tau_i\psi_\sigma(\mathbf{x})\right)^2.
\end{eqnarray}
As mentioned above, we will consider three types of disorder: 1) random chemical potential,
2) random gauge potential and 3) random mass.
Its general representation adopted from \co{Refs.~\onlinecite{Nersesyan1995NPB,Stauber2005PRB,Wang2011PRB}}, is:
\begin{eqnarray}
H_{\mathrm{dis}}= \nu_m \int d^2\mathbf{x}\psi^\dagger(\mathbf{x})M\psi(\mathbf{x})A(\mathbf{x}).
\end{eqnarray}
Here $M=\tau_0$ is the random chemical potential, $M=\tau_1$ and $M=\tau_3$ the random gauge potential
(two components), and $M=\tau_2$ the random mass disorders.  The field $A(\mathbf{x})$ represents a quenched,
Gauss-white potential determined by $\langle A(\mathbf{x}) \rangle =0 $, while
$\langle A(\mathbf{x}) A(\mathbf{x}') \rangle = n_0\delta(\mathbf{x}-\mathbf{x}') $, \de{where $n_0$ is the impurity (defects) concentration}.
\textcolor{black}{The impurity scattering rate we quantify by
$\tau^{-1} = n_0\nu^2_m/t$, which will be measured by $\Lambda_E=t\Lambda^2$}
(for more details see Supplementary materials (SM) \cite{Supple_materials}).
 \de{In general, a complete analysis should contain}
 \co{all possible
fermion bilinears including those
appearing due to interaction effects. However, we focus here}
on the suppression of the topological phases by disorder. Therefore for the sake of simplicity we restrict ourself
to the effect of the aforementioned types of disorder separately.

\begin{figure}[t]
\centering
\includegraphics[width=3.5in]{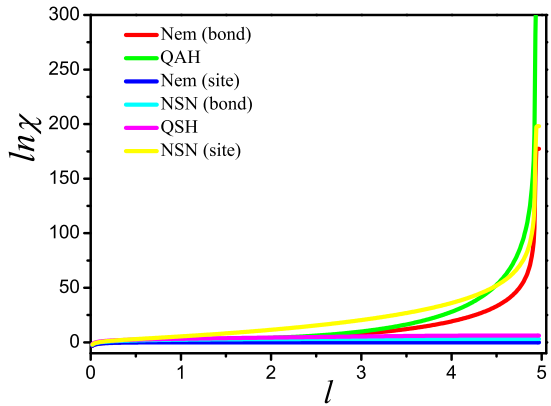}
\vspace{-1.05cm}
\caption{(Color online) Susceptibilities to particle-hole phases as functions
of the RG flow parameter $l$ \textcolor{black}{by approaching the QAH FP $(g^*_0,g^*_-,g^*_2)/g_+ =(0,-3.73,7.46)$
for checkerboard lattice.}}\label{Fig_chi_checkerborad}
\end{figure}

\emph{RG analysis and fixed points --}
Within the Wilsonian renormalization group theory \cite{Wilson1975RMP,Polchinski9210046,Shankar1994RMP,Huh2008PRB,She2010PRB,Wang2013NJP,She2015PRB,
Vafek2010PRB_2,Vafek2014PRL}, we derive the flow equations by integrating out the fields in the
momentum shell $e^{-l}\Lambda<k<\Lambda$, with $l>0$ the running scale, integrating over all
frequencies at the same time. The disorder contributes both to the fermion self-energy renormalization
and vertices renormalization. As a result, we obtain a system of flow equations for the coupling constants $g_i$,
disorder potentials $\nu_m$ and $t$ in the form \cite{Fradkin2009PRL,Vafek2012PRB,Vafek2014PRB}:
\begin{eqnarray}
\frac{d g_i}{d l} &=& \sum_{jk} A_{ijk} g_j g_k + \sum_{j} B_{ij} \frac{n_0 \nu_m^2}{\pi t^2}g_j, \label{Eq_flow_g_i} \nonumber \\
\frac{d \nu_m}{d l} &=& \left(D_0 \nu_m + t \sum_i D_i  g_i \right) \frac{n_0 \nu_m^2}{\pi t^2} , \label{Eq_flow}\\
\frac{d t}{d l} &=& - C \frac{n_0 \nu_m^2}{\pi t^2} t, \nonumber \label{Eq_flow_t}
\end{eqnarray}
where the coefficients  $A_{ijk}$, $B_{ij}$, $C$, $D_0$ and $D_{i}$  are provided in the SM \cite{Supple_materials}.
The  fixed points (FPs) are subsequently determined from the numerical analysis of the flow equations
\co{Eq.~(\ref{Eq_flow}}). Solving the flow equations in the clean limit \cite{Vafek2014PRB,Vafek2012PRB}
leads to three fixed points $(g^*_0,g^*_-,g^*_2)=(0,-3.73,7.46)g_+$, $(g^*_0,g^*_-,g^*_2)=(0,3.73,7.46)g_+$ and $(g^*_0,g^*_-,g^*_2)=(0,0,-1.09)g_+$, where  $g_{\pm}=(g_3\pm g_1)/2$. The first two fixed points correspond to the QAH order
while the last one to QSH.

Under the influence of the disorder the fixed points move in the space of \co{the} coupling
constants. \textcolor{black}{The evolution of the FP $(0,-3.73, 7.46)g_+$ and FP $(g^*_0,g^*_-,g^*_2)=(0,0,-1.09)g_+$ with
increasing the strength of the bare random chemical potential are shown in the inset figures of
Fig.~\ref{Fig_Phase_diagram_chemical}. They gradually change with
the increase
of the bare values of disorder and
saturate with an intermediate plateau, \textcolor{black}{designated \co{as} QAH (QSH) and QAH-II (QSH-II,III) FPs in Fig.~\ref{Fig_Phase_diagram_chemical} (here and below for easy identification we label the fixed points by the corresponding ground states QAH and QSH)}.
We found that other evolutions are also possible (for disorders of the  random mass and
random gauge potential types see \cite{Supple_materials})}.

\begin{figure}[t]
\centering
\includegraphics[width=2.2in]{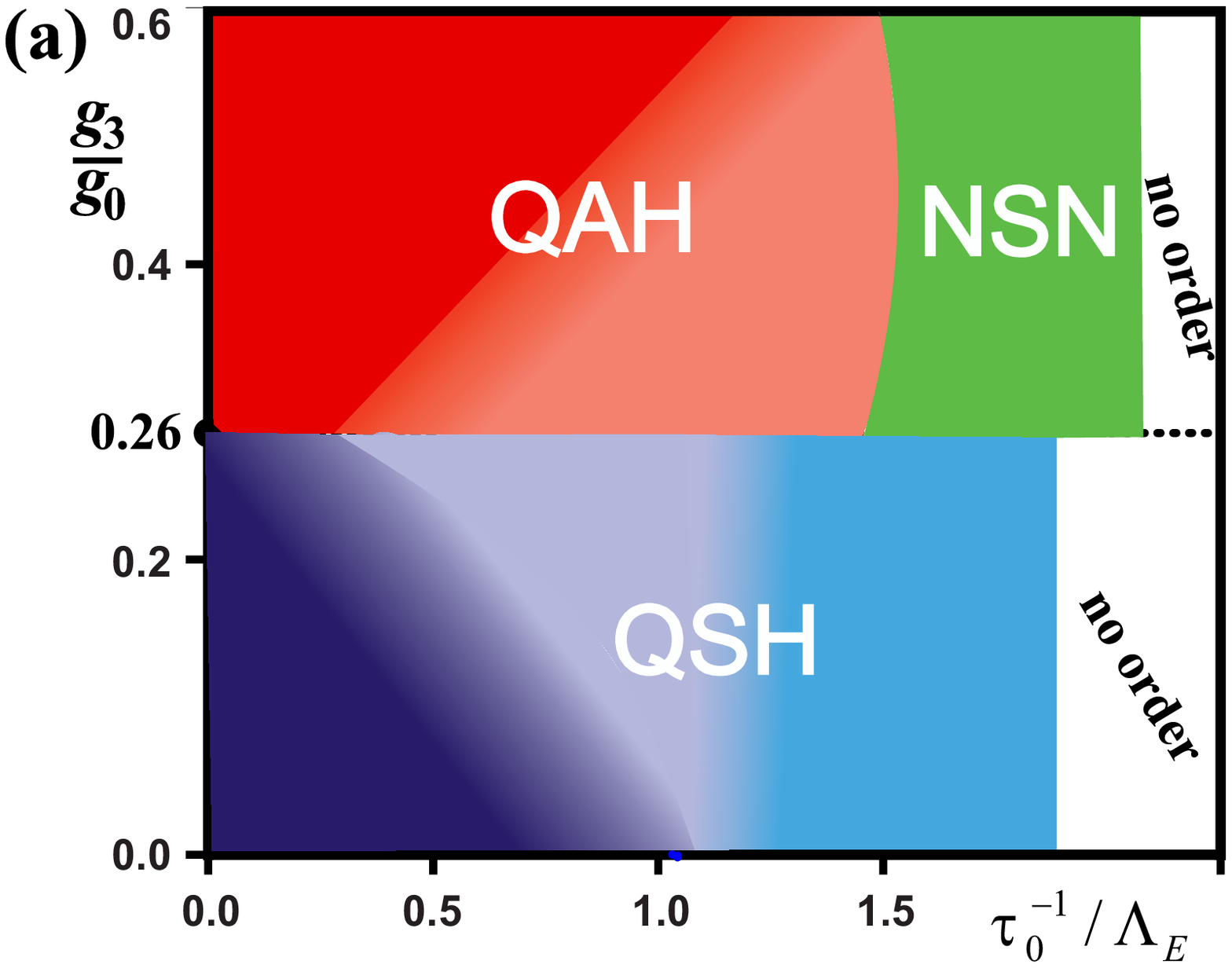}\vspace{0.16cm}
\includegraphics[width=2.2in]{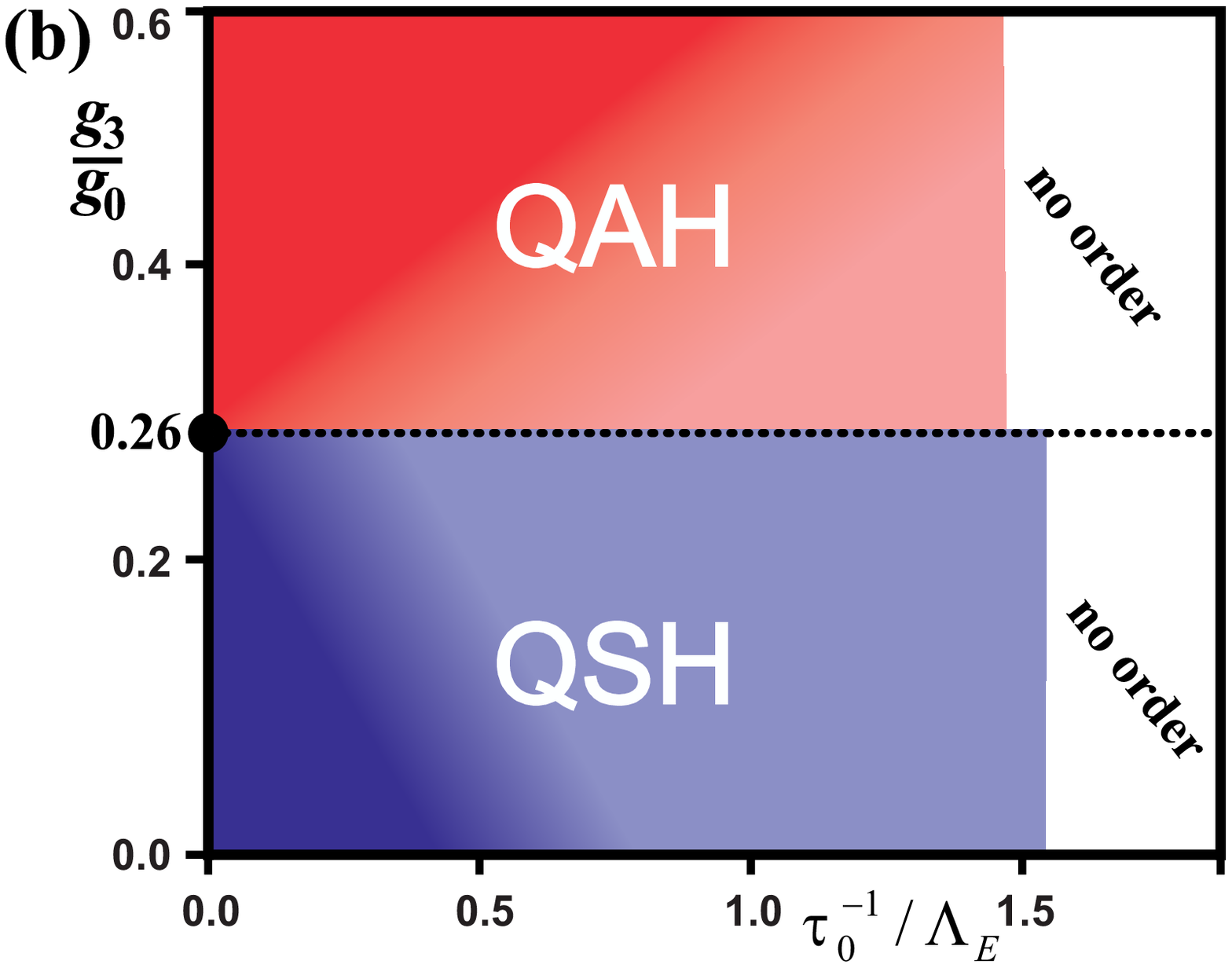}
\vspace{-0.2cm}
\caption{(Color online) Schematic phase diagram as a function of disorder and interaction strength
$g_3/g_0$ for the checkerboard lattice. (a): random chemical potential and (b): random mass.
\textcolor{black}{The change of the color from dark to light is deduced from the evolution
of the FPs as shown in Fig.~\ref{Fig_Phase_diagram_chemical}.}}\label{Fig_Phase_diagram_checkerboard}
\end{figure}

\begin{figure}
\centering
\includegraphics[width=2.2in]{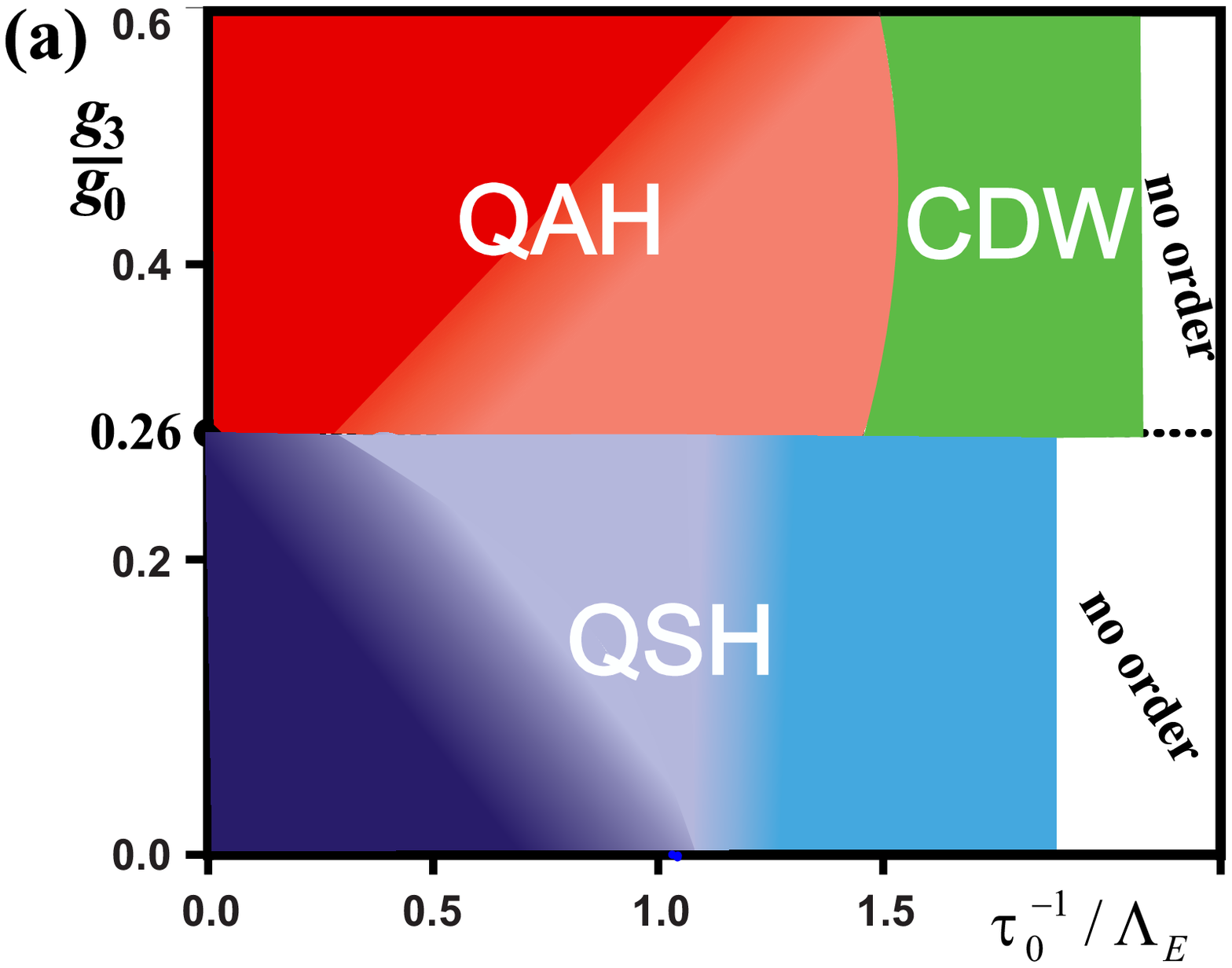}\vspace{0.16cm}
\includegraphics[width=2.2in]{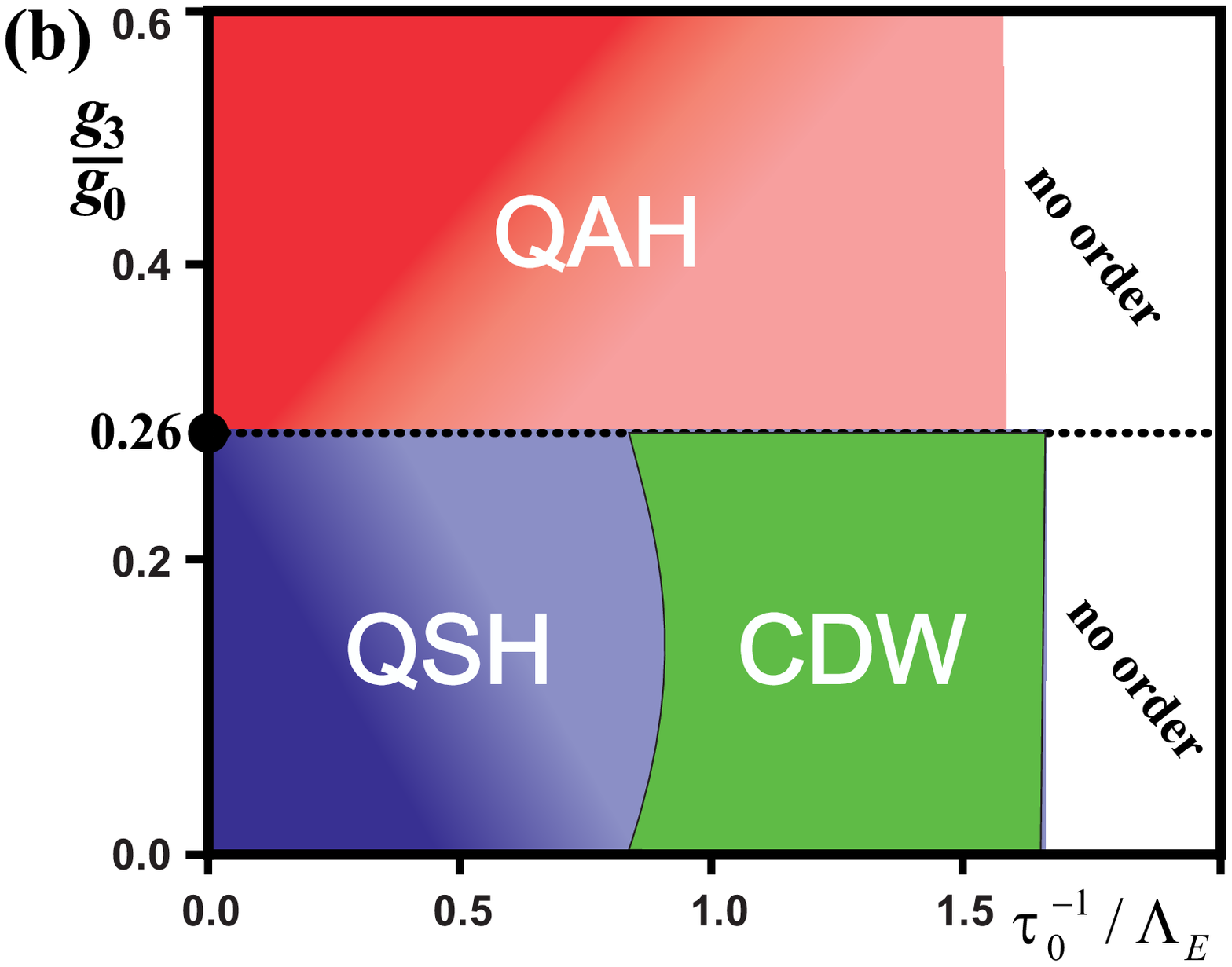}
\vspace{-0.2cm}
\caption{(Color online) Schematic phase diagrams for the honeycomb lattice for bare couplings $g_1(l=0)=g_2(l=0)=0$
in the presence of (a): random chemical potential and (b): random mass. \textcolor{black}{The change of the color
from dark to light is deduced from the evolution of the FPs as shown in Fig.~\ref{Fig_Phase_diagram_chemical}.}}\label{Fig_Phase_diagram_honeycomb}
\end{figure}

\emph{Susceptibilities and phase diagrams --}
To find the phases that are
realized at the FPs, we \co{next} solve the flow equations for the order parameters $\Delta_i$ corresponding
to the different long-range  orders allowed by the symmetries of the corresponding crystal structures. All possible
\de{order parameters for the checkerboard and honeycomb lattices} are listed in the SM \cite{Supple_materials}.  In particular,
 a finite expectation value of the time-reversal symmetry-breaking order parameter $\Delta_{\mathrm{QAH}}$ corresponds to a QAH phase with quantized Hall conductivity, as it can be shown, in the clean limit and at the mean-field level, by integrating the Berry curvature in the full BZ \cite{Fradkin2009PRL,Supple_materials}. Similarly, the ground state with order parameter $\Delta_{\mathrm{QSH}}$ signals the onset of the spin-rotation symmetry-breaking QSH phase, which is characterized by the spin Chern number $(C^{\uparrow}-C^{\downarrow})/2$~\cite{Kirtschig2015}.

Employing the relation $\delta\chi_i(l)=-\frac{\partial^2 \delta f}{\partial \Delta_i(0)
\partial\Delta_i^*(0)}$ \cite{Vafek2012PRB,Vafek2014PRB,Nelson1975PRB}, where $f$ is the free energy,
we can obtain the corresponding susceptibilities approaching the FPs. One finds
that near the RG scale $l^*_c$, where the couplings $g_i(l)$ diverge, $\chi_i(l)\sim(l^*_c-l)^{-\gamma}$.
For instance, their behavior as a function of the RG flow around the QAH FP is
depicted in Fig.~\ref{Fig_chi_checkerborad} and others are provided in SM \cite{Supple_materials}.
Therefore the ground state may be obtained as the state
characterized by the
 susceptibility with the strongest
divergence or by comparison of the corresponding critical indexes $\gamma_i$ \cite{Vafek2012PRB,Vafek2014PRB}.
We checked that both ways give the same results.

Using this procedure we determine the resulting ground state  as a function of the ratio of
the bare interaction strength $g_3/g_0$ and the disorder strength. The phase diagram for the
checkerboard lattice is shown in Fig.~\ref{Fig_Phase_diagram_checkerboard}. Please note that
the boarder lines are drawn schematically and are the matter
of further investigations. In the clean limit the $g_3/g_0<0.26$ corresponds to QSH state,
while $g_3/g_0>0.26$ to QAH state \cite{Vafek2014PRB}. Considering the QAH state in the
presence of the chemical potential disorder,  one sees that it is changed at certain value
of disorder by the spin nematic (NSN) site order. Further increase of disorder
potential leads to the non-ordered state. In contrast, the QSH state is suppressed
by disorder without changes to intermediate phases. For the random mass potential
we found no intermediate phases.

\textcolor{black}{To further understand the possible consequences of the fixed point evolution,
we subsequently determine the phase diagram with an effective $T$-dependence, which is linked to the
transformation $T=T_0e^{-l}$ \cite{Huh2008PRB,She2015PRB}. }As critical
temperature we use the value $T_c=T_0e^{-l^*_c}$, \textcolor{black}{the results are presented
in Fig.~\ref{Fig_Phase_diagram_chemical}.} Considering the effective critical temperature as a
function of the random chemical potential disorder, one notes a considerable change of
the $T_c$ slope at the QAH $\to$ NSN transition in Fig.~\ref{Fig_Phase_diagram_chemical}(a).
Surprisingly, the slopes also considerably change with evolution of the fixed points
\textcolor{black}{within the same phase.
This comes from
the fast crossovers from one FP to another.}
An example of such a slope change is provided by the evolution between the \textcolor{black}{
QSH-II and QSH-III FPs} in Fig.~\ref{Fig_Phase_diagram_chemical}(b).
\textcolor{black}{The evolution of the FP is gradual and one does not see
any characteristic features on $T_c$ \cite{Supple_materials}.
}
\jb{The situation of random mass and random gauge potential is detailed in the SM \cite{Supple_materials}.}
Before summarizing the results we have to note that the scattering rate $\tau^{-1}(l)$ is  strongly   renormalized in 2D together with the interaction \cite{Supple_materials}. Therefore  the experimentally relevant values of the impurity scattering rate are not the bare $\tau_0$ but  $\tau^{-1}(l_c^*)$ at the characteristic energy of the instability. The comparison of  values of the scattering rate $\tau^{-1}(l_c^*)$ necessary to suppress the effective critical temperature twice with the effective critical temperature \co{$T_{c}^{0}$} in the clean limit is \jb{summarized} in Table~\ref{table_phase}.
\textcolor{black}{As one can see from the table, the critical impurity scattering rates \co{for a complete} suppression of the topological phases are of the order of the critical temperature} \co{in the clean limit.
By considering that the latter corresponds to the dynamically generated gap, we can conclude that, in perfect analogy with non-interacting topological insulating states ~\cite{Halperin1982PRB}, the stability of interaction-driven topological insulators is relatively immune to non-magnetic impurities.}

\begin{table}[t]
\centerline{
\begin{tabular}{| c |r|r|r|r|}
\hline
& QAH & QSH & NSN & CDW \\
\hline
C & $\tau^{-1}(l^*)\sim T^0_c$ & $\tau^{-1}(l^*)\sim T^0_c$ & $\tau^{-1}(l^*) \gg T^0_c$ & $\tau^{-1}(l^*) \gg T^0_c$  \\
\hline
M & $\tau^{-1}(l^*)\sim T^0_c$ & $\tau^{-1}(l^*)\sim T^0_c$ & - & -  \\
\hline
G & $\tau^{-1}(l^*)\sim T^0_c$ & $\tau^{-1}(l^*)\sim T^0_c$ & - & -  \\
\hline
\end{tabular}
}
\caption{Stability of the phases against different types of disorder:  chemical potential (C), random mass (M) and random gauge potential (G). Here $\tau^{-1}(l^*_{c0})$ is taken at the energy of the first instability  in the clean limit.
}\label{table_phase}
\end{table}

\emph{Bilayer graphene --}
We subsequently generalize our analysis to the honeycomb lattice model for bilayer graphene.
At the two inequivalent ${\bf K}$ and $\bf {K^{\prime}}$ points of the Brillouin zone, the low-energy bands touch parabolically
\cite{McCann2006PRL,Nilsson2008PRB,Vafek2010PRB,Vafek2010PRB_2,Venderbos2016}, and realize a two-valley QBCP system
with an effective Hamiltonian
$H=\sum_{\mathbf{k}}\psi^\dagger_{\mathbf{k}\sigma}\mathcal{H}_0\psi_{\mathbf{k}\sigma}$
(the noninteracting Hamiltonian and other information are provided in SM \cite{Supple_materials}).
The effective action \cite{Vafek2010PRB,Vafek2012PRB} is similar to the one for the checkerboard lattice and
provided explicitly in the SM \cite{Supple_materials}. Despite the effective theory for the checkerboard and the bilayer honeycomb
lattice are similar, the latter exhibits many more instabilities \cite{Vafek2012PRB}
due to the additional valley degree of freedom. In order to fully understand the leading instabilities for
the bilayer honeycomb lattice, we calculate the susceptibilities for all
18 possible orderings \cite{Vafek2012PRB,Vafek2010PRB,Lemonik2010PRB,Haldane1988PRL,
Nandkishore2010PRB,Varma2013PRB,MacDonald2008PRB,Nandkishore2010PRL,Mudry2007PRL,Throckmorton2012PRB,Lemonik2012PRB,
Scherer2012PRB,Neto2009RMP,Vafek2010PRB_2,Kharitonov2012PRB}.
The resulting disorder phase diagrams for the bilayer honeycomb lattice are shown in Fig.~\ref{Fig_Phase_diagram_honeycomb}.
The temperature-dependent phase diagrams at bare $g_1(l=0)=g_2(l=0)=0$ are qualitatively similar to the ones of
the checkerboard lattice, but with the NSN (site) and QSH phases in Fig.~\ref{Fig_Phase_diagram_chemical}
replaced by a charge-density wave (CDW) phase  \cite{Supple_materials}.

We have analyzed the fate of the weak-coupling interaction-driven topological
insulators phases realized in 2D Fermi systems with a QBCP, under the influence of disorder.
By means of the RG approach and unbiasedly studying the fermion-interacting couplings and disorders,
we build the coupled flow equations of the fermion-interacting couplings and disorder strength.
We established that the different types of disorder
\co{generally suppress the critical temperature at which the interaction-driven topological states set in.}
In particular cases, \co{strong} disorder can even induce phase transition from a topological to
a non-topologically ordered state. Disorder in interaction-driven topological systems thus gives rise to
a distinct set of phenomena that can be looked for and studied experimentally. Moreover, the response to
disorder might be used as an experimental signature that a material is actually in a, so-far unobserved,
interaction-driven topologically insulating state of matter.

\begin{acknowledgments}
We acknowledge J. M. Murray and O. Vafek for useful correspondence \textcolor{black}{ and  C. Fulga for helpful discussions}.
J.W. is supported by the National Natural Science Foundation of China under Grant 11504360,
the China Postdoctoral Science Foundation under Grants 2015T80655 and 2014M560510, the Fundamental
Research Funds for the Central Universities, and the Program of Study Abroad for Young Scholar sponsored by
China Scholarship Council. C.O. acknowledges the financial support of the Future and Emerging Technologies Programme for Research of the European Commission under FET-Open grant number: 618083 (CNTQC), and the Deutsche Forschungsgemeinschaft under Grant No. OR 404/1-1.
D.V.E. and J.v.d.B would like to acknowledge the financial support provided by the German Research Foundation
(Deutsche Forschungsgemeinschaft) through the program DFG-Russia, BR4064/5-1. J.v.d.B is also supported by SFB
1143 of the Deutsche Forschungsgemeinschaft. \textcolor{black}{D.V.E would also acknowledge the VW foundation for partial financial support.}
\end{acknowledgments}


%
%
%
%
%
%
%
%
%




\newpage

\appendix
\onecolumngrid
\section{Supplementary Materials for: "The fate of interaction-driven topological insulators under disorder"}

%
%

\maketitle
\section{Model and Effective theory}\label{Sec_model}
\de{{\it Lattice model} -- The minimal model on the checkerboard lattice is:
\begin{eqnarray}
H=\sum_{ij} (- t_{ij}) c^\dagger_i c_j + V \sum c^\dagger_i c^\dagger_j c_j c_i,
\end{eqnarray}
\co{where $t_{ij}$ is the hopping amplitude between sites $i$ and $j$ while $V>0$ is the nearest-neighbor repulsion. Moreover, $t_{ij}=t,t^{\prime},t^{\prime \prime}$, respectively for nearest neighbors, and next-nearest neighbors connected or not by a diagonal bond}.
\co{Since} the checkerboard lattice has two sublattices  $A$ an $B$, 
it is useful to introduce a spinor   $\Psi^\dagger_i = (c_{iA}^\dagger,c_{iB}^\dagger)$.  Then the free particle Hamiltonian reads}:
\co{
\begin{equation}
H_0 = \sum_{l=0}^3 \sum_{\mathbf{k}\sigma}\epsilon_{l}(\mathbf{k}) \psi_{\mathbf{k}\sigma}^\dagger \tau_l \psi_{\mathbf{k}\sigma}
\label{eq.TB.checkerboard}
\end{equation}
where  $\epsilon_{0}({\bf k})=-(t^{\prime} + t^{\prime \prime}) \left(\cos{k_x} + \cos{k_y} \right)$, $\epsilon_{1}({\bf k})=  4 t \cos{(\frac{kx}{2})} \cos{(\frac{ky}{2})}$, and $\epsilon_{3}({\bf k})=-(t^{\prime}-t^{\prime \prime}) \left(\cos{k_x} - \cos{k_y} \right)$.
}

\de{\it Low energy sector--} The noninteracting Hamiltonian for the checkerboard lattice \de{in the low energy sector} \co{can be obtained expanding the tight-binding model near the corner of the Brillouin zone, {\it i.e.} at the $M=\left(\pi, \pi \right)$ point, and } is given by \cite{Fradkin2009PRL}
\begin{eqnarray}
H_0=\sum_{|\mathbf{k}|<\Lambda}\sum_{\sigma=\uparrow\downarrow}\psi^\dagger_{\mathbf{k}\sigma}
\mathcal{H}_0(\mathbf{k})\psi_{\mathbf{k}\sigma},
\end{eqnarray}
where
\begin{eqnarray}
\textcolor{black}{\mathcal{H}_0(\mathbf{k})
=t_I\mathbf{k}^2I+2t_xk_xk_y\tau_1+t_z(k^2_x-k^2_y)\tau_3.}
\end{eqnarray}
\co{The parameters of the continuum Hamiltonian are related to the hopping amplitudes by $t_x=t/2$, $t_I=(t^{\prime}+t^{\prime \prime})/2$, and $t_z=(t^{\prime} - t^{\prime \prime})/2$.}

The primary interacting part is written as \cite{Fradkin2009PRL,Vafek2012PRB,Vafek2014PRB},
\begin{eqnarray}
H_{\mathrm{int}}=\sum_{i}\frac{2\pi}{m}g_i\int d^2\mathbf{x}\left(\sum_{\sigma=\uparrow\downarrow}\psi^\dagger_\sigma(\mathbf{x})
\tau_i\psi_\sigma(\mathbf{x})\right)^2
\end{eqnarray}
with $\tau_i$ being the Pauli matrices, \textcolor{black}{which is allowed by the symmetries \cite{Fradkin2009PRL,Vafek2012PRB,Vafek2014PRB,
Wen2008PRB,Fradkin2008PRB}.} The eigenvalues of $\mathcal{H}_0(\mathbf{k})$ are derived as \cite{Fradkin2009PRL,Vafek2014PRB}
\begin{eqnarray}
E^{\pm}_{\mathbf{k}}=\frac{\mathbf{k}^2}{\sqrt{2}m}\left[\lambda\pm\sqrt{\cos^2\eta\cos^2\theta_{\mathbf{k}}
+\sin^2\eta\sin^2\theta_{\mathbf{k}}}\right],
\end{eqnarray}
where $m=\frac{1}{\sqrt{2(t^2_x+t^2_z)}}$, $\lambda=\frac{t_I}{\sqrt{t^2_x+t^2_z}}$,
$\cos\eta=\frac{t_z}{\sqrt{t^2_x+t^2_z}}$, and $\sin\eta=\frac{t_x}{\sqrt{t^2_x+t^2_z}}$ \cite{Vafek2014PRB}.
Here $\psi_{\mathbf{k}\sigma}$ has two components, which in the case of a checkerboard lattice correspond
to sublattices A and B, above equation describes one upward and one downward dispersing band at
$|t_I|<\min(|t_x|,|t_z|)$ \cite{Fradkin2009PRL,Vafek2014PRB}. The Hamiltonian possesses two touching parabolically
at $\mathbf{k}=0$ and is invariant under the $C_4$ point group and time-reversal symmetry \cite{Fradkin2009PRL,Vafek2014PRB}.

We here stress that the disorder $A(\mathbf{x})$ is a quenched,
Gaussian white noise potential defined by the following correlation functions
\begin{eqnarray}
\left\langle A(\mathbf{x})\right\rangle=0;\hspace{0.5cm}
\left\langle A(\mathbf{x}_1)A(\mathbf{x}_2)\right\rangle=n_0\delta^{2}(\mathbf{x}_1-\mathbf{x}_2),
\end{eqnarray}
the dimensionless parameter $n_0$ represents the concentration of impurity.

\begin{figure}
\centering
\includegraphics[width=5.6in]{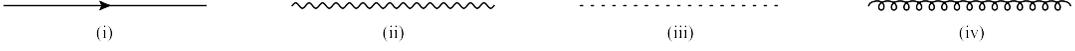}
\vspace{-0.3cm}
\caption{\textcolor{black}{Free propagators for (i) fermion, (ii) fermionic interaction, (iii) disorder, and (iv) source field.}}\label{Fig_propagators}
\end{figure}

Without lost of generality and also in order to compare with the according results in Ref. \cite{Vafek2014PRB}, we here primarily
concentrate on the case in the limit of particle-hole symmetry $(\lambda=0)$ and rotational invariance
$(\eta=\frac{\pi}{4})$.  After the Fourier transformations and involving above analysis,
we finally obtain the effective action in the presence of disorder,
\begin{eqnarray}
S_{\mathrm{eff}}&=&\int^{+\infty}_{-\infty}\frac{d\omega}{2\pi}\int\frac{d^2\mathbf{k}}{(2\pi)^2}\sum_{\sigma=\uparrow\downarrow}
\psi^\dagger_{\sigma}(\omega,\mathbf{k})[-i\omega+2tk_xk_y\tau_1+t(k^2_x-k^2_y)\tau_3]\psi_{\sigma}(\omega,\mathbf{k})
+\frac{2\pi}{m}\sum^{3}_{i=0}g_i\int^{+\infty}_{-\infty}\frac{d\omega_1d\omega_2d\omega_3}{(2\pi)^3}\nonumber\\
&&\times\int^{\Lambda}\frac{d^2\mathbf{k}_1d^2\mathbf{k}_2d^2\mathbf{k}_3}{(2\pi)^6}\sum_{\sigma,\sigma'=\uparrow\downarrow}
\psi^\dagger_\sigma(\omega_1,\mathbf{k}_1)\tau_i\psi_\sigma(\omega_2,\mathbf{k}_2)\psi^\dagger_{\sigma'}(\omega_3,\mathbf{k}_3)
\tau_i\psi_{\sigma'}(\omega_1+\omega_2-\omega_3,\mathbf{k}_1+\mathbf{k}_2-\mathbf{k}_3)\nonumber\\
&&+\nu_m\int^{+\infty}_{-\infty}\frac{d\omega}{2\pi}\int \frac{d^2\mathbf{k}d^2\mathbf{k'}}{(2\pi)^4}\psi^\dagger(\mathbf{k},\omega)
M\psi(\mathbf{k'},\omega)A(\mathbf{k-k'}).\label{Eq_effective_action}
\end{eqnarray}
with $m=1/(2t)$ and $M=\tau_0$ being the random chemical potential, $M=\tau_1$ and $M=\tau_3$ the random gauge potential
(two components), and $M=\tau_2$ the random mass \cite{Stauber2005PRB,Wang2011PRB}, respectively. Here the parameter $\nu_m$
measures the strength of a single impurity \textcolor{black}{and the corresponding impurity scattering rate can be expressed as
$\tau^{-1}\sim n_0\nu^2_m/t$, which will be measured by $\Lambda_E=t\Lambda^2$ with $t$ rescaled by $\hbar^2/2m$.}
The free propagators are represented in the Fig.~\ref{Fig_propagators}.

\section{Coupled flow equations and fixed points}\label{Sec_RG_analysis}

\begin{figure}
\centering
\includegraphics[width=6.2in]{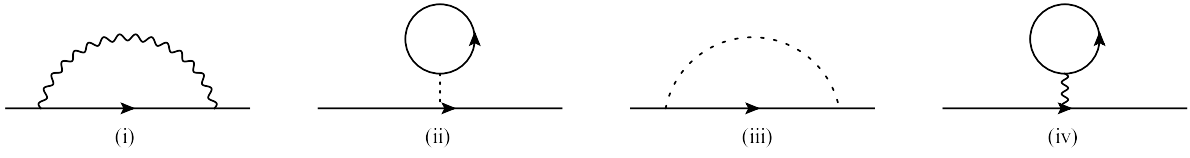}
\vspace{-0.3cm}
\caption{One-loop corrections to the fermion propagator.}\label{Fig_fermion_propagator_correction}
\end{figure}

\begin{figure}
\centering
\includegraphics[width=4.3in]{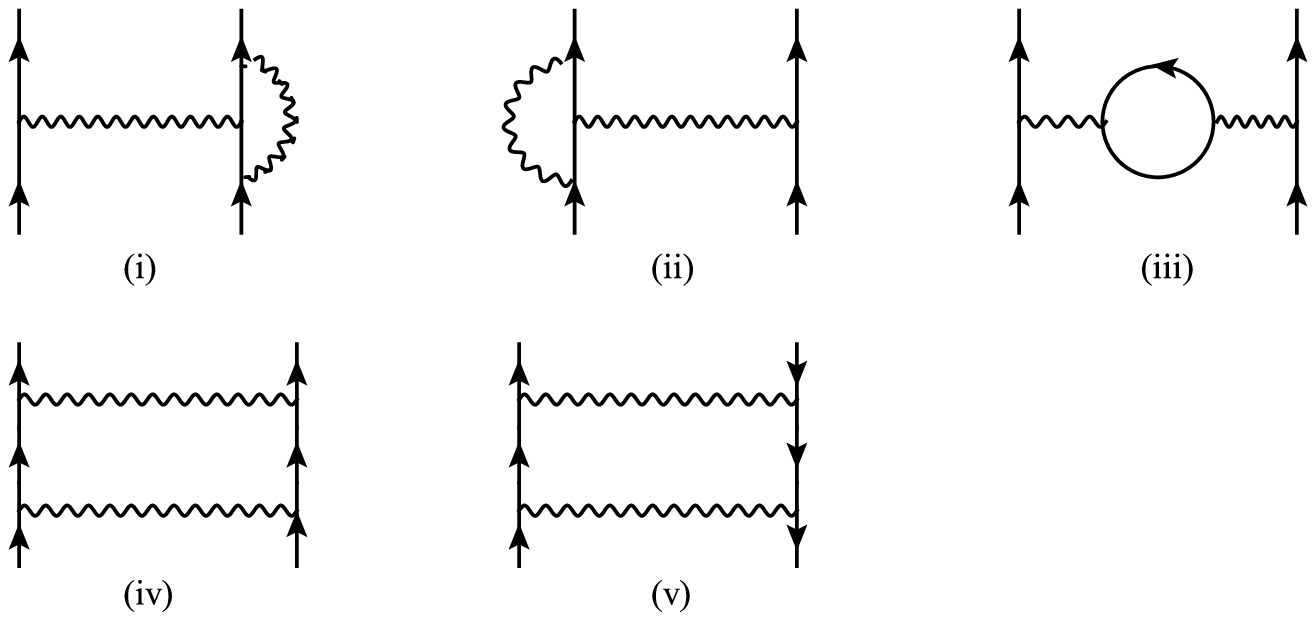}
\vspace{-0.1cm}
\caption{One-loop corrections to the fermion interacting couplings due to fermionic  interactions.}\label{Fig_fermion_interaction_correction_clean}
\end{figure}

\begin{figure}
\centering
\includegraphics[width=5.0in]{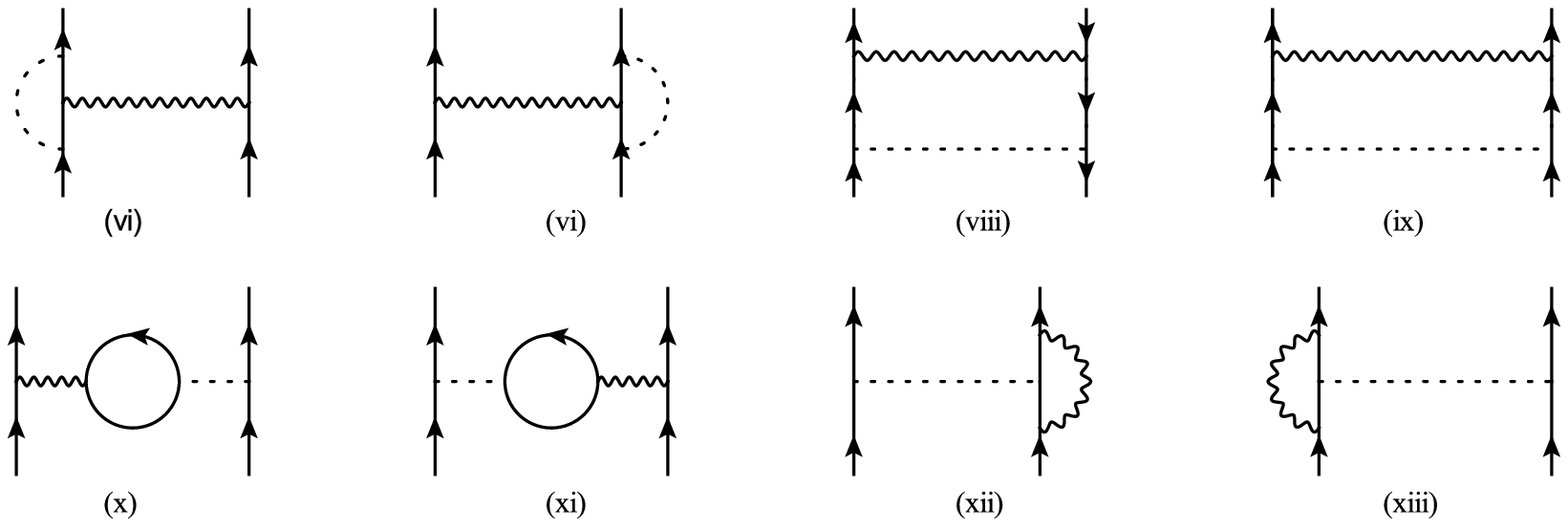}
\vspace{-0.1cm}
\caption{One-loop corrections to the fermion interacting couplings due to the fermionic interactions
and disorder effects.}\label{Fig_fermion_interaction_correction_disorder}
\end{figure}

By including the disorder corrections and considering the RG theory \cite{Shankar1994RMP,Huh2008PRB,She2010PRB}, we obtain the revised re-scaling transformation as given in the main text. In the presence of disorder, the fermions receive self-energy corrections from the fermion-disorder interaction as shown in Fig.~\ref{Fig_fermion_propagator_correction}. In addition, the one-loop corrections
to the fermion interacting couplings and the fermion-disorder vertex in presence of different sorts of disorders
as presented in Fig.~\ref{Fig_fermion_interaction_correction_clean} and Fig.~\ref{Fig_fermion_interaction_correction_disorder}.
After calculating the one-loop corrections paralleling the steps in Refs. \cite{Huh2008PRB,She2010PRB,Wang2011PRB,Shankar1994RMP},
we derive the coupled flow equations for all parameters. Denoting $g_+=\frac{g_3+g_1}{2}$ and $g_-=\frac{g_3-g_1}{2}$ \cite{Vafek2012PRB,Vafek2014PRB}, we obtain the reduced flow equations for all parameters, as listed in the following. In the presence of random chemical potential with $M=\tau_0$, the coupled flow equations are
\begin{eqnarray}
\frac{dt}{dl}&=&-\left(\frac{n_0 \nu^2_m}{4\pi t^2}\right)t,\\
\frac{d g_0}{dl}&=&\left(-4g_+-\frac{3n_0 \nu^2_m}{2\pi t^2}\right)g_0,\\
\frac{d g_+}{dl}&=&\Bigl[-(g_0-g_+)^2-(g_2-g_+)^2-6g^2_+-\frac{n_0 \nu^2_m}{2\pi t^2}g_+\Bigr],\\
\frac{d g_2}{dl}&=&\left[4(g_0g_2-g^2_2-g^2_-+g^2_+-3g_2g_+)-\frac{2n_0 \nu^2_m}{\pi t^2}g_2\right],\\
\frac{d g_-}{dl}&=&\Big[2g_-(g_0-3g_2-2g_+)-\frac{n_0 \nu^2_m}{2\pi t^2}g_-\Bigr],\\
\frac{d \nu_m}{dl}&=&\frac{[n_0 \nu^2_m-8\pi t(g_0+g_2+2g_+)]}{4\pi t^2}\nu_m.
\end{eqnarray}
In the presence of random gauge potential with $M=\tau_{1,3}$ whose flow equations are the same, the coupled flow equations look like:
for both $M=\tau_{1}$ and $M=\tau_{3}$,
\begin{eqnarray}
\frac{dt}{dl}&=&-\left(\frac{n_0 \nu^2_m}{4\pi t^2}\right)t,\\
\frac{d g_0}{dl}&=&-4g_+g_0+\frac{n_0 \nu^2_m}{2\pi t^2}g_0,\\
\frac{d g_+}{dl}&=&\Bigl[-(g_0-g_+)^2-(g_2-g_+)^2-6g^2_+\Bigr],\\
\frac{d g_2}{dl}&=&\left[4(g_0g_2-g^2_2-g^2_-+g^2_+-3g_2g_+)+\frac{n_0 \nu^2_m}{2\pi t^2}g_2\right],\\
\frac{d g_-}{dl}&=&\Big[2g_-(g_0-3g_2-2g_+)\Bigr],\\
\frac{d \nu_m}{dl}&=&-\left(\frac{n_0 \nu^2_m}{4\pi t^2}\right)\nu_m.
\end{eqnarray}
Finally, the coupled flow equations in the presence of random mass with $M=\tau_2$ read:
\begin{eqnarray}
\frac{dt}{dl}&=&-\left(\frac{n_0 \nu^2_m}{4\pi t^2}\right)t,\\
\frac{d g_0}{dl}&=&-4g_+g_0,\\
\frac{d g_+}{dl}&=&\Bigl[-(g_0-g_+)^2-(g_2-g_+)^2-6g^2_+-\frac{n_0 \nu^2_m}{2\pi t^2}g_+\Bigr],\\
\frac{d g_2}{dl}&=&\left[4(g_0g_2-g^2_2-g^2_-+g^2_+-3g_2g_+)+\frac{n_0 \nu^2_m}{2\pi t^2}g_2\right],\\
\frac{d g_-}{dl}&=&\Big[2g_-(g_0-3g_2-2g_+)-\frac{n_0 \nu^2_m}{2\pi t^2}g_-\Bigr],\\
\frac{d \nu_m}{dl}&=&\frac{[8\pi t(g_0-2g_++g_2)-3n_0 \nu^2_m]}{4\pi t^2}\nu_m.
\end{eqnarray}

\begin{figure}
\centering
\includegraphics[width=2.6in]{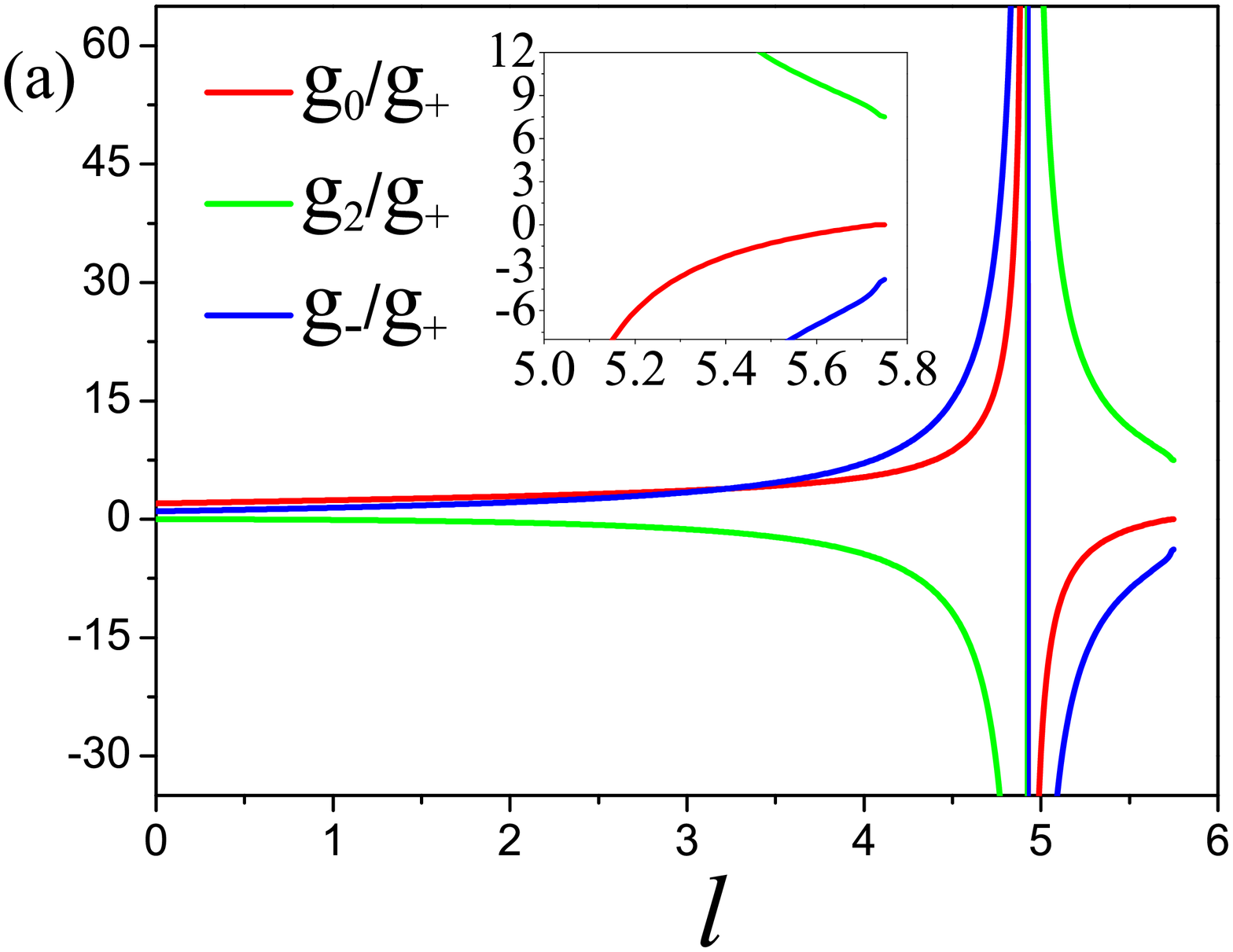}\hspace{-1.36cm}
\includegraphics[width=2.6in]{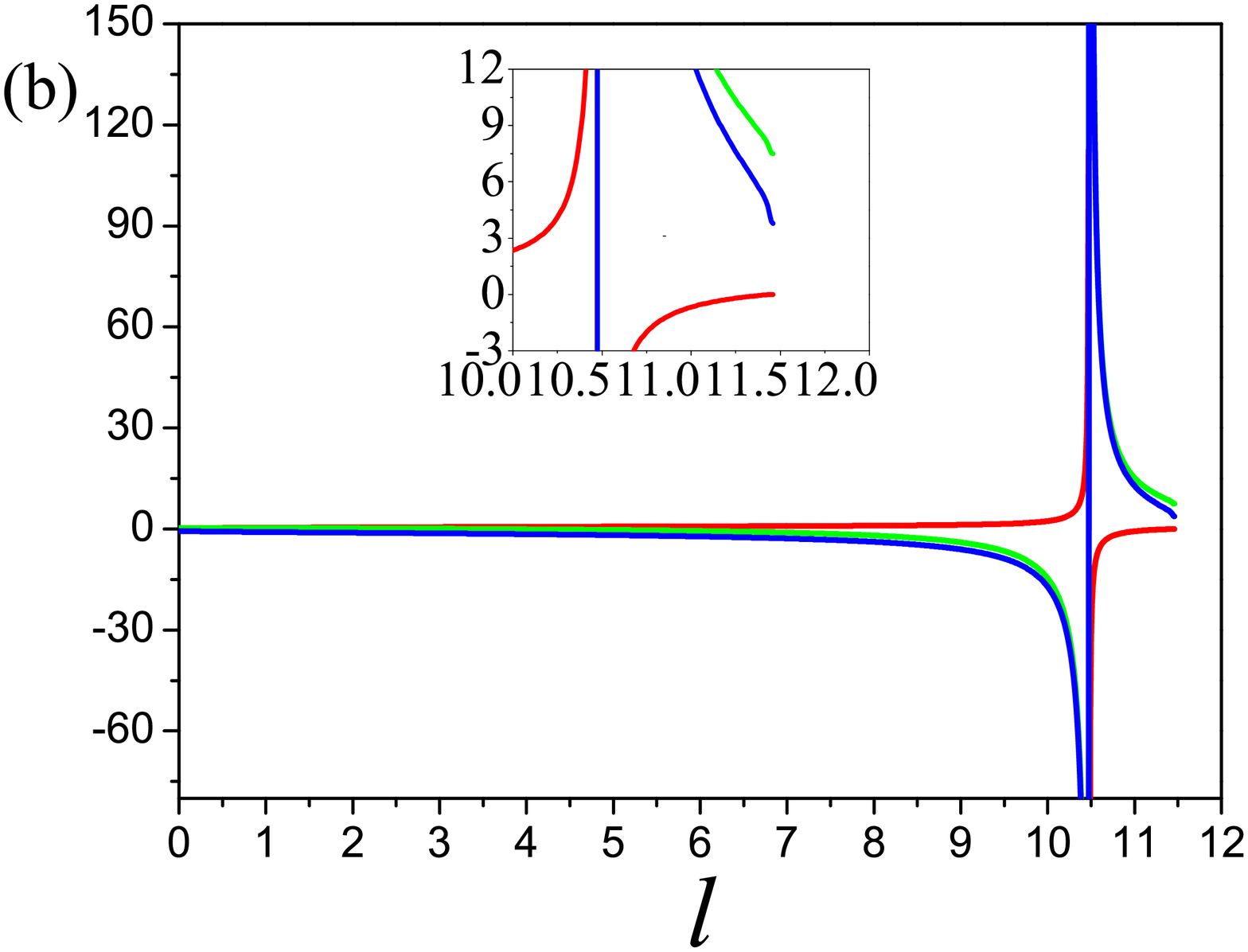}\hspace{-1.36cm}
\includegraphics[width=2.6in]{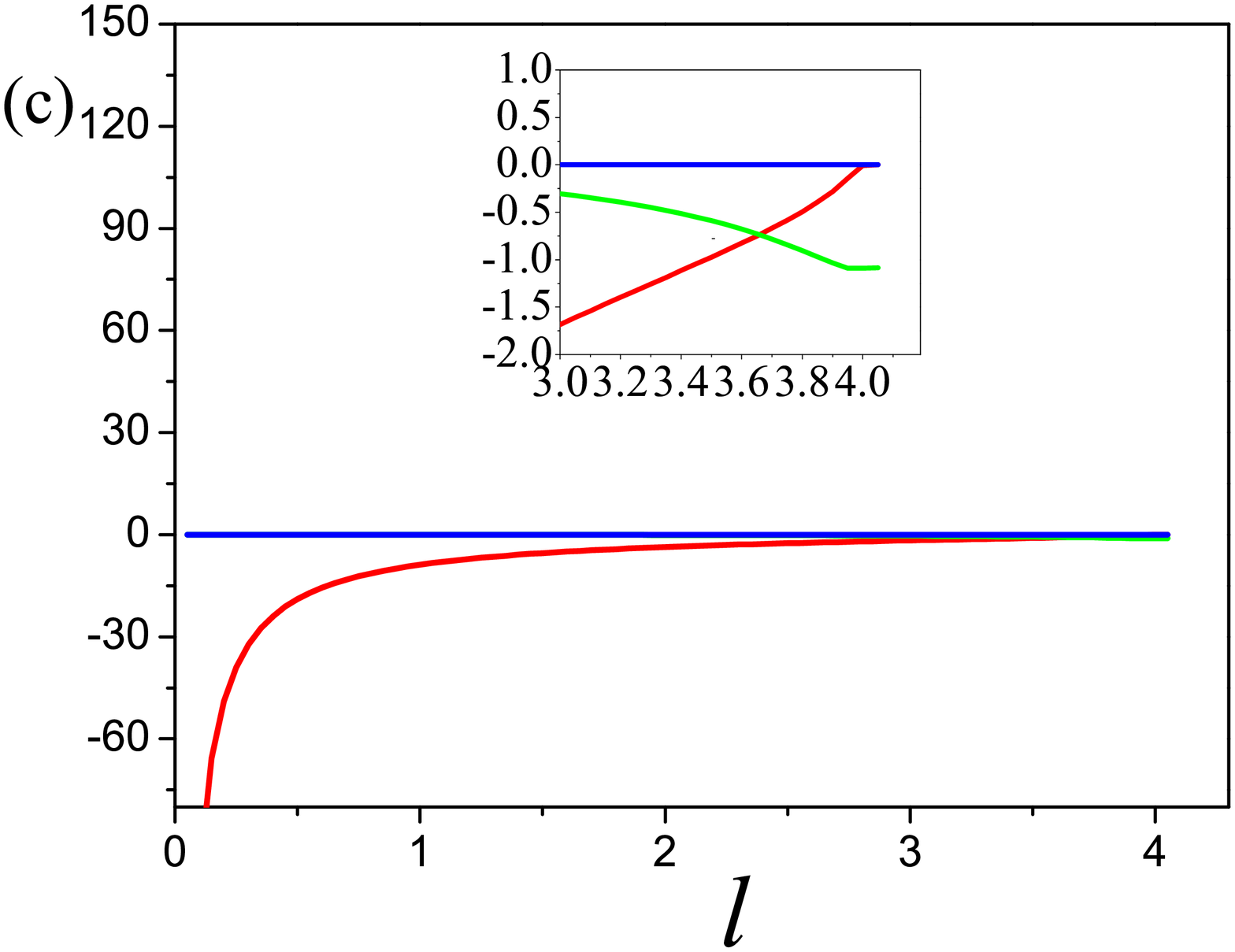}
\vspace{-0.36cm}
\caption{Flows of $g_0/g_+$, $g_2/g_+$ and $g_-/g_+$ in clean limit at some representatively initial values.
The $(g_0,g_2,g_-)/g_+$ finally towards three fixed points: (a) $(g^*_0,g^*_-,g^*_2)/g_+=(0,-3.73,7.46)$; (b) $(g^*_0,g^*_-,g^*_2)/g_+=(0,3.73,7.46)$ and (c) $(g^*_0,g^*_-,g^*_2)/g_+=(0,0,-1.09)g^+$.
Inset: the enlarged regime for the fixed point.}\label{Fig_FP_clean_limit}
\end{figure}

Performing numerical calculations of
above coupled flow equations, we get the fixed points. The trajectories towards to the fixed points in clean limit have already
studied by Murray and Vafek \cite{Vafek2014PRB}. For completeness, we provide the corresponding trajectories
in Fig.~\ref{Fig_FP_clean_limit}. Next, we consider the fixed points in the presence of different types of disorder.
The evolution of fixed points in the presence of random chemical potential,
and random mass are presented in Fig.~\ref{Fig_M0_dis} and Fig.~\ref{Fig_M2_dis} respectively. In distinction to
the other two types of disorders, we find that both the QAH fixed point $(g^*_0,g^*_-,g^*_2)=(0,-3.73,7.46)$ and
the QSH fixed point $(g^*_0,g^*_-,g^*_2)=(0,0,-1.09)$ are robust against
 the random gauge potential and do not evolve with increase of the disorder. By these reasons we do not show them here.

\begin{figure}
\centering
\includegraphics[width=4.3in]{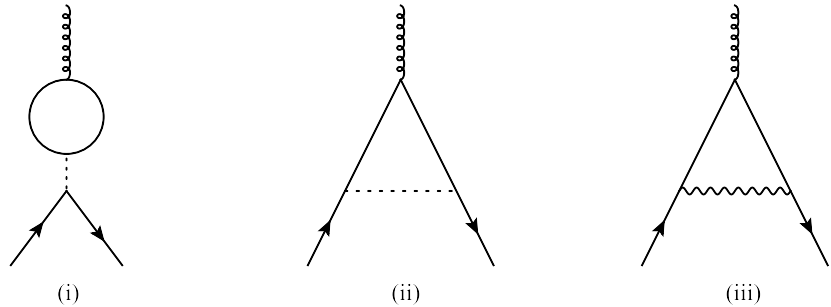}
\vspace{-0.1cm}
\caption{One-loop corrections to the fermion-source terms $\Delta^{c,s}$.}\label{Fig_fermion_source_correction}
\end{figure}

\section{Mean field order parameter }\label{Sec_order_parameters}

In order to investigate possible types of symmetry breaking, we collect both the charge and spin source terms into the action
\cite{Vafek2012PRB,Vafek2014PRB}:
\begin{eqnarray}
S_\Delta
=\int d\tau\int d^2\mathbf{x}\sum^3_{i=0}\left(\Delta^{c}_i\psi^\dagger \mathcal{M}^{c}_i\psi
+\vec{\Delta}^{s}_i\cdot\psi^\dagger \mathbf{\mathcal{M}}^{s}_i\psi\right).
\end{eqnarray}
The matrices $\mathcal{M}^{c,s}$ define the various fermion bilinears in charge and spin channels \cite{Vafek2014PRB}.
One-loop corrections to the fermion-source terms $\Delta^{c,s}$ can be derived by computing the diagrams
in Fig.~\ref{Fig_fermion_source_correction}. For the charge channel, the matrixes
\begin{equation} \mathcal{M}^c_1=\tau_1, \mbox{~~}\mathcal{M}^c_2=\tau_2,  \mbox{ and  }  \mathcal{M}^c_3=\tau_3
\label{eq.order.parameter.continuum}
\end{equation}
correspond to the nematic (bond), QAH, and nematic (site),
respectively \cite{Vafek2014PRB}.  Besides, for the spin channel, the matrixes $\mathcal{M}^s_0=1\vec{s}$,
$\mathcal{M}^s_1=\tau_1\vec{s}$, $\mathcal{M}^s_2=\tau_2\vec{s}$, and $\mathcal{M}^s_3=\tau_3\vec{s}$ refer to the
FM, NSN (bond), QSH, and NSN (site), respectively \cite{Vafek2014PRB}. Since the susceptibilities of $\mathcal{M}^{c,s}_0$
are independent of the energy scale which are not primary instabilities, we will only pay significance
to the instabilities for cases $\mathcal{M}^{c,s}_{1,2,3}$ in the following text.



\de{\paragraph*{Chern number--}
The topological invariant in the insulating phase with finite expectation value of $\mathcal{M}^c_2$ can be computed using a relation between the Chern number and the $2 \pi$ Berry flux carried by a symmetry-protected quadratic band crossing point.
The latter corresponds to the topological charge of the $\hat{d}$ vector vortex with $\hat{d}=\left\{2 k_x k_y, k_x^2-k_y^2\right\}$.
Using a gauge fixing procedure~\cite{Kirtschig2015}, one can indeed show that the Chern number acquired by breaking time-reversal symmetry can be related to the topological charge $W$ as $C=W/2$~\cite{Kirtschig2015}. This equivalence allows to obtain the Chern number for a quadratic band crossing point even though in this model the momentum space cannot be one-point compactified to a unit sphere ${\cal S}_2$. And indeed, the Chern number computed in this way corresponds precisely to the Chern number obtained at the mean field level using the lattice formulation~\cite{Fradkin2009PRL}.
A similar analysis can be performed for the QSH phase using that in each spin channel carries an opposite Hall conductivity and hence Chern number.}

\de{Another approach of calculation of the Chern number is based on  mapping of the continuous model onto a tight binding model. Then the Chern number is given by the integration over the Brillouin zone (BZ):
\begin{equation}
N_{\mbox{Chern}} = \frac{1}{4\pi}\int_{\mathrm{BZ}} dk_x dk_y \left(\mathbf{n} \cdot \left[ \frac{\partial  \mathbf{n}}{\partial k_x} \times \frac{\partial \mathbf{n} }{\partial k_y} \right]   \right).
\label{eq.chern}
\end{equation}
For the introduced tight binding model Eq. (\ref{eq.TB.checkerboard}) and the mean-field order parameter of the QAH state $\mathcal{M}_2^c= \sin(\frac{k_x}{2})\sin(\frac{k_y}{2})\tau_2$ the  vector $\mathbf{n} = \mathbf{d}/d$ with
\begin{equation}
\mathbf{d} = t\left(4 \cos\frac{k_x}{2}\cos\frac{k_y}{2}, \frac{\Delta_{QAH}}{2t} (\sin\frac{k_x}{2}\sin\frac{k_y}{2}), -(\cos k_x-\cos k_y)  \right).
\end{equation}
Calculating the integral Eq. (\ref{eq.chern}) we find that the Chern number  coincides with the discussed above continuous model $N_{\mbox{Chern}}=- sign (\Delta_{QAH})$.}

\section{Susceptibilities in the presence of disorder}\label{Sec_susceptibilities}

\begin{figure}
\centering
\includegraphics[width=3.6in]{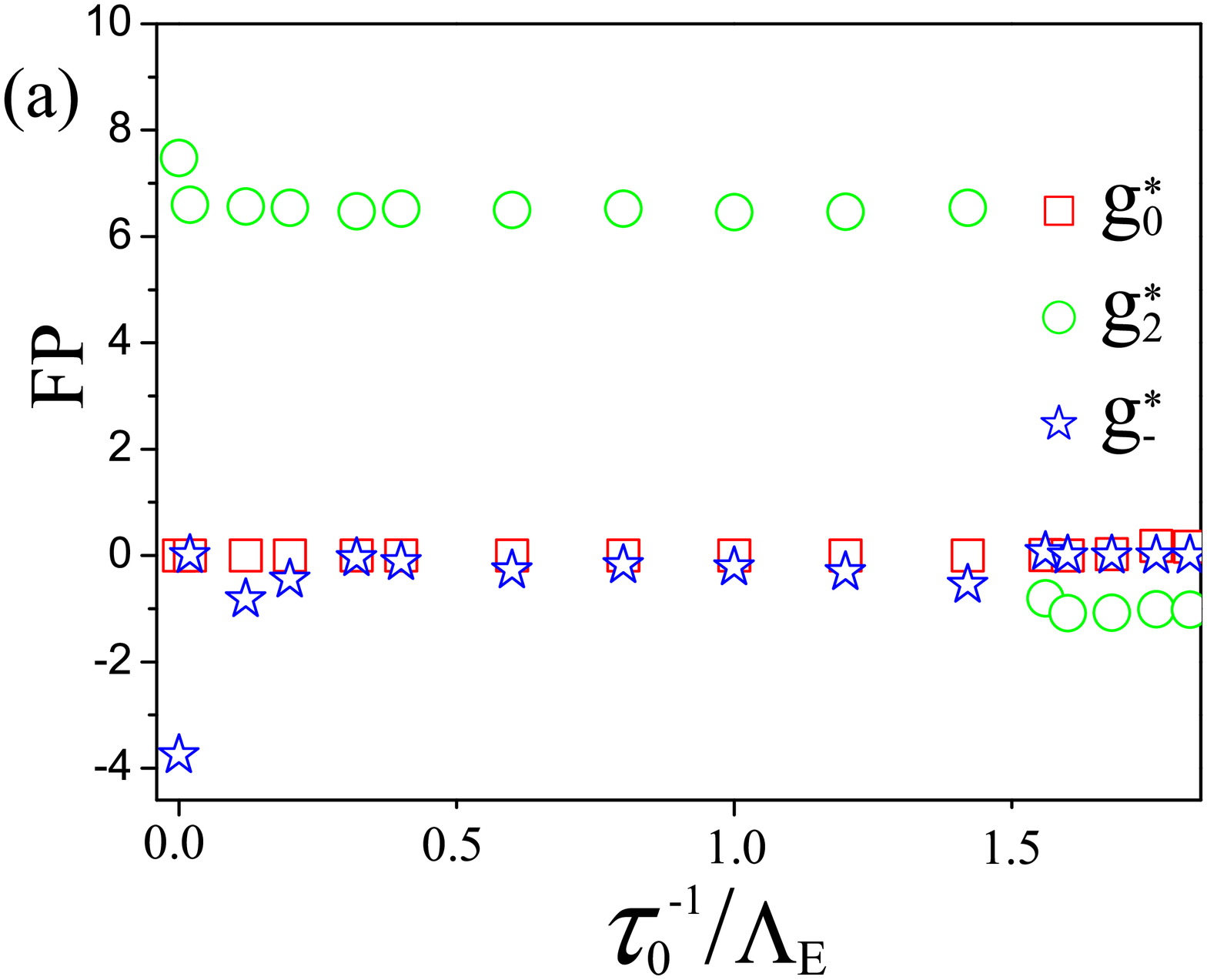}\hspace{-0.80cm}
\includegraphics[width=3.6in]{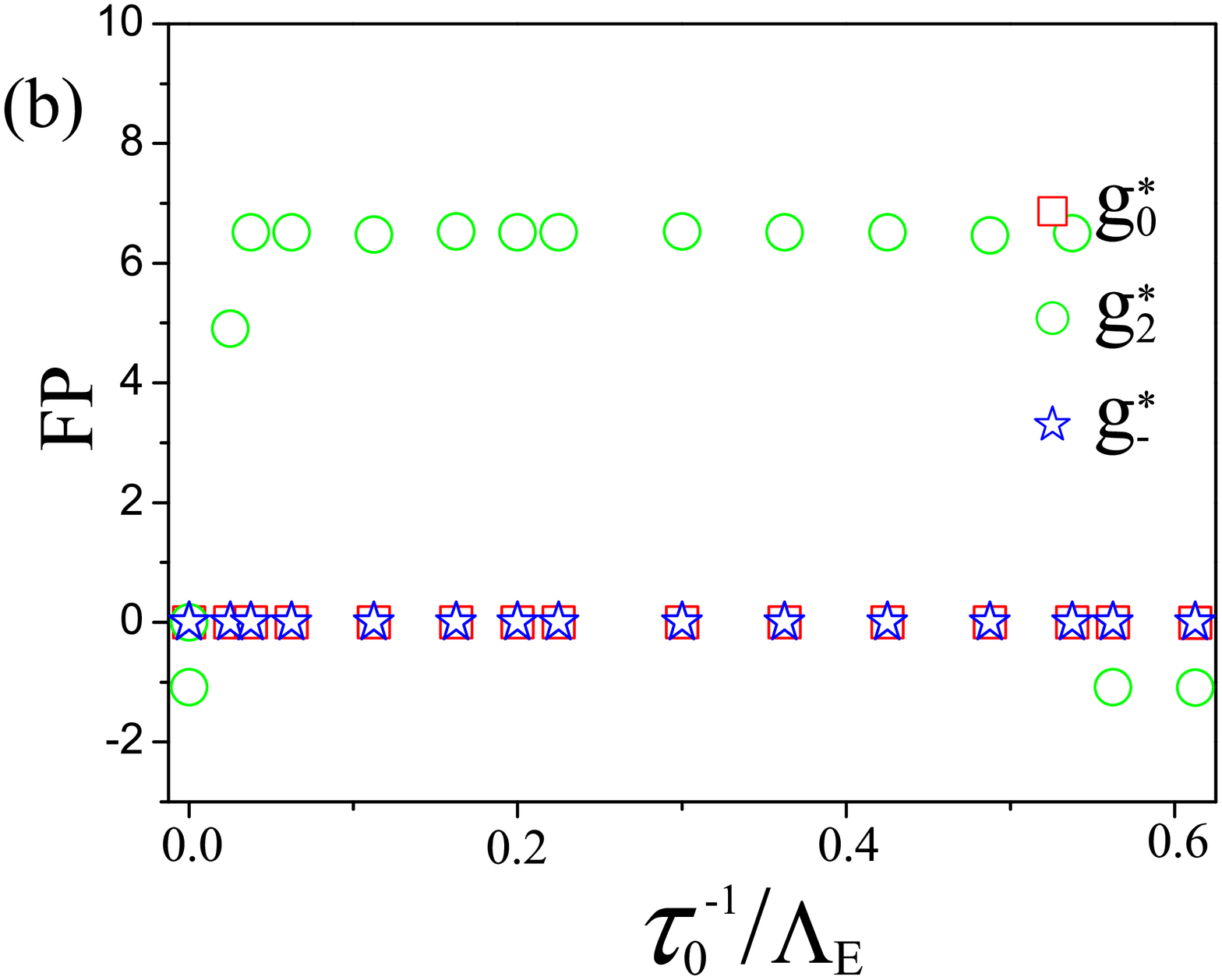}
\vspace{-0.36cm}
\caption{(Color online) Evolution of the fixed points $(g^*_0,g^*_-,g^*_2)$ for the checkerboard lattice
in the presence of random chemical potential around (a): the QAH fixed point $(0,-3.73,7.46)$
and (b): the QSH fixed point $(0,0,-1.09)$ at some representatively initial values of
disorder strength of the random chemical potential. FP represents the fixed points
and The parameter $\tau^{-1}\sim n_0\nu^2_m/t$ designates the impurity scattering rate and
$\Lambda_E=t\Lambda^2$ with $t$ rescaled by $\hbar^2/2m$.}\label{Fig_M0_dis}
\end{figure}

\begin{figure}
\centering
\includegraphics[width=3.6in]{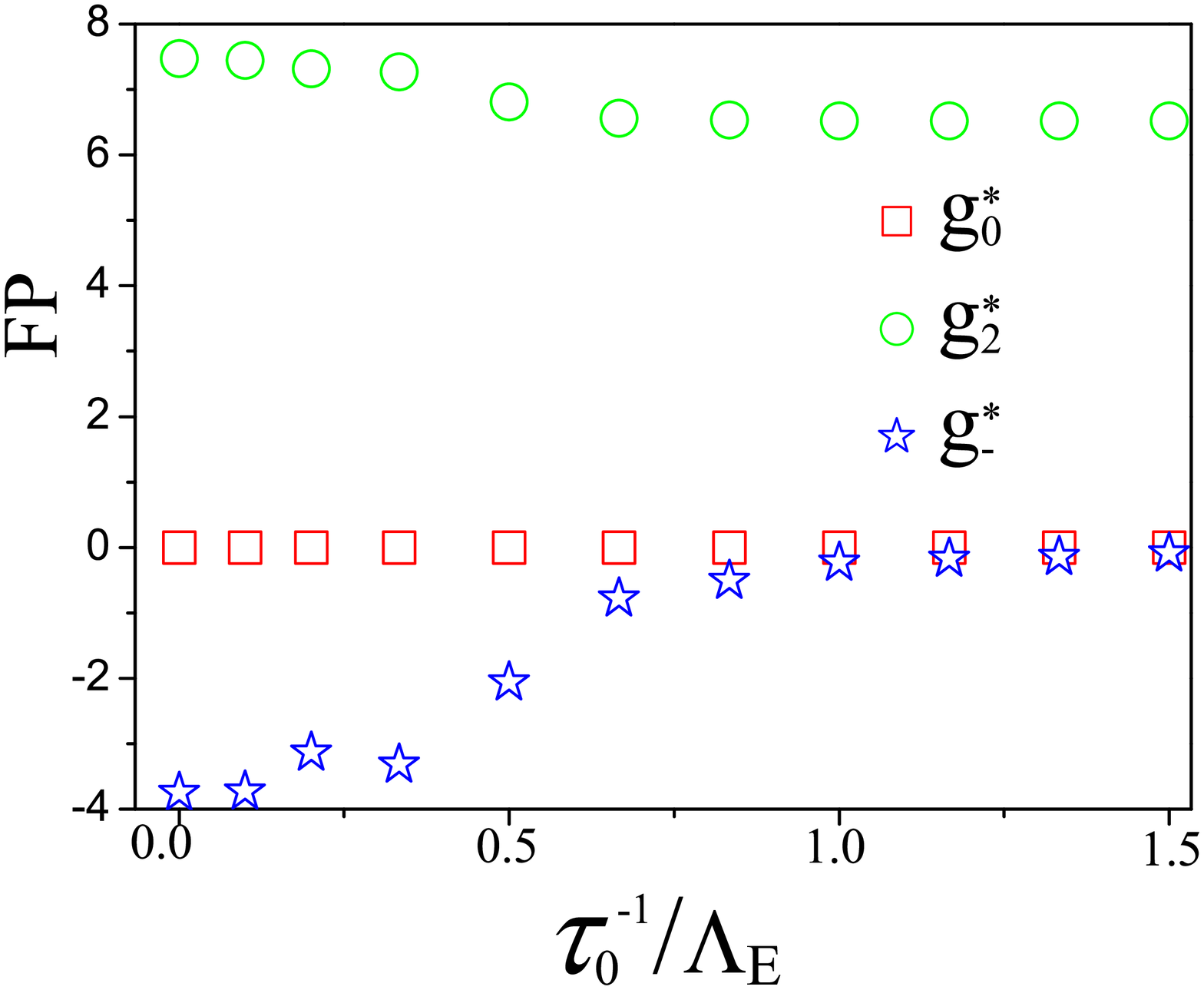}\hspace{-0.80cm}
\includegraphics[width=3.6in]{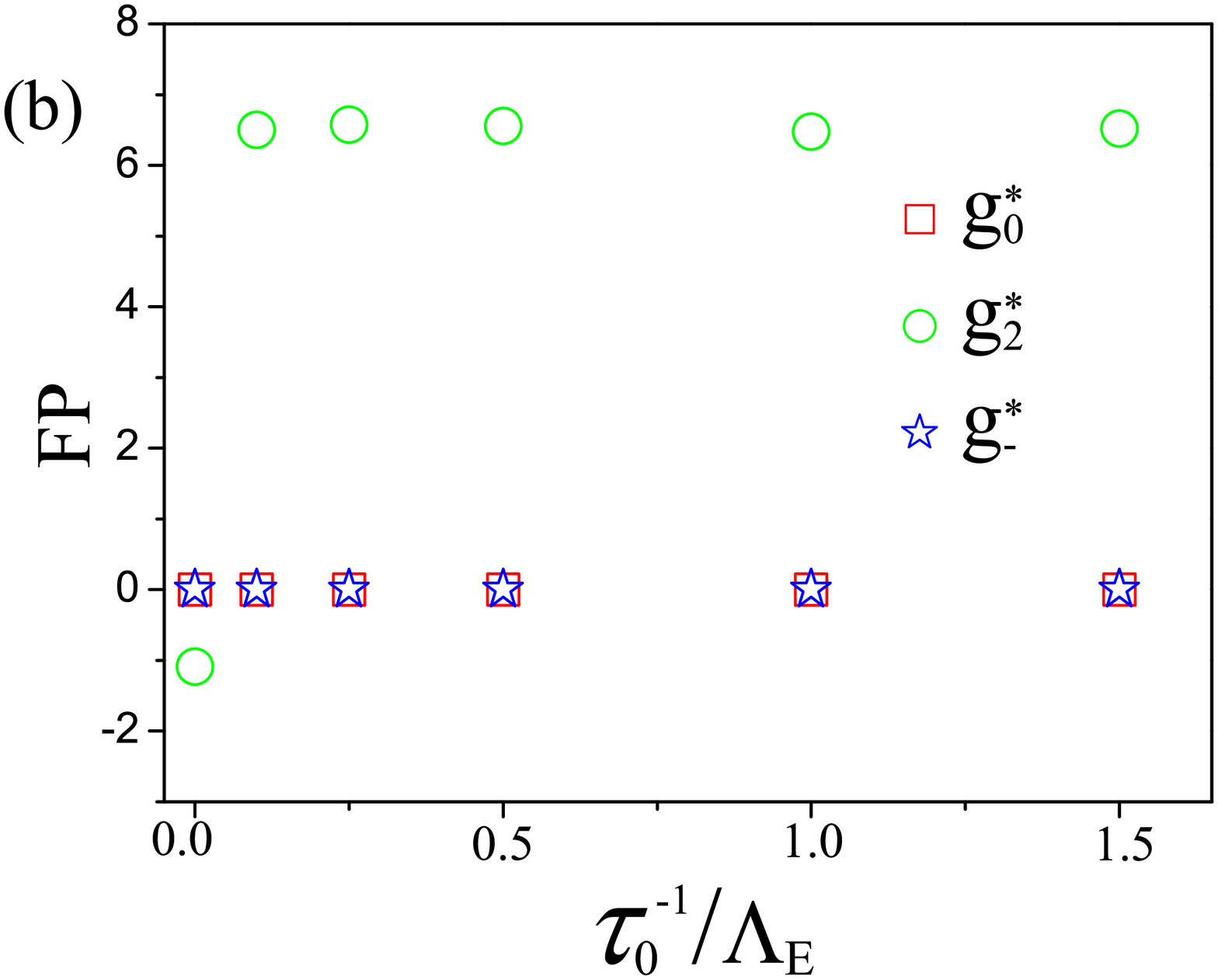}
\vspace{-0.36cm}
\caption{(Color online) \textcolor{black}{Evolution of the fixed points $(g^*_0,g^*_-,g^*_2)$ for the checkerboard lattice
in the presence of random mass around (a): the QAH fixed point $(0,-3.73,7.46)g_+$
and (b): the QSH fixed point $(0,0,-1.09)g_+$ at some representatively initial values of
disorder strength of the random chemical potential. FP represents the fixed points
and The parameter $\tau^{-1}\sim n_0\nu^2_m/t$ designates the impurity scattering rate and
$\Lambda_E=t\Lambda^2$ with $t$ rescaled by $\hbar^2/2m$.}}\label{Fig_M2_dis}
\end{figure}

\begin{figure}
\centering
\includegraphics[width=3.6in]{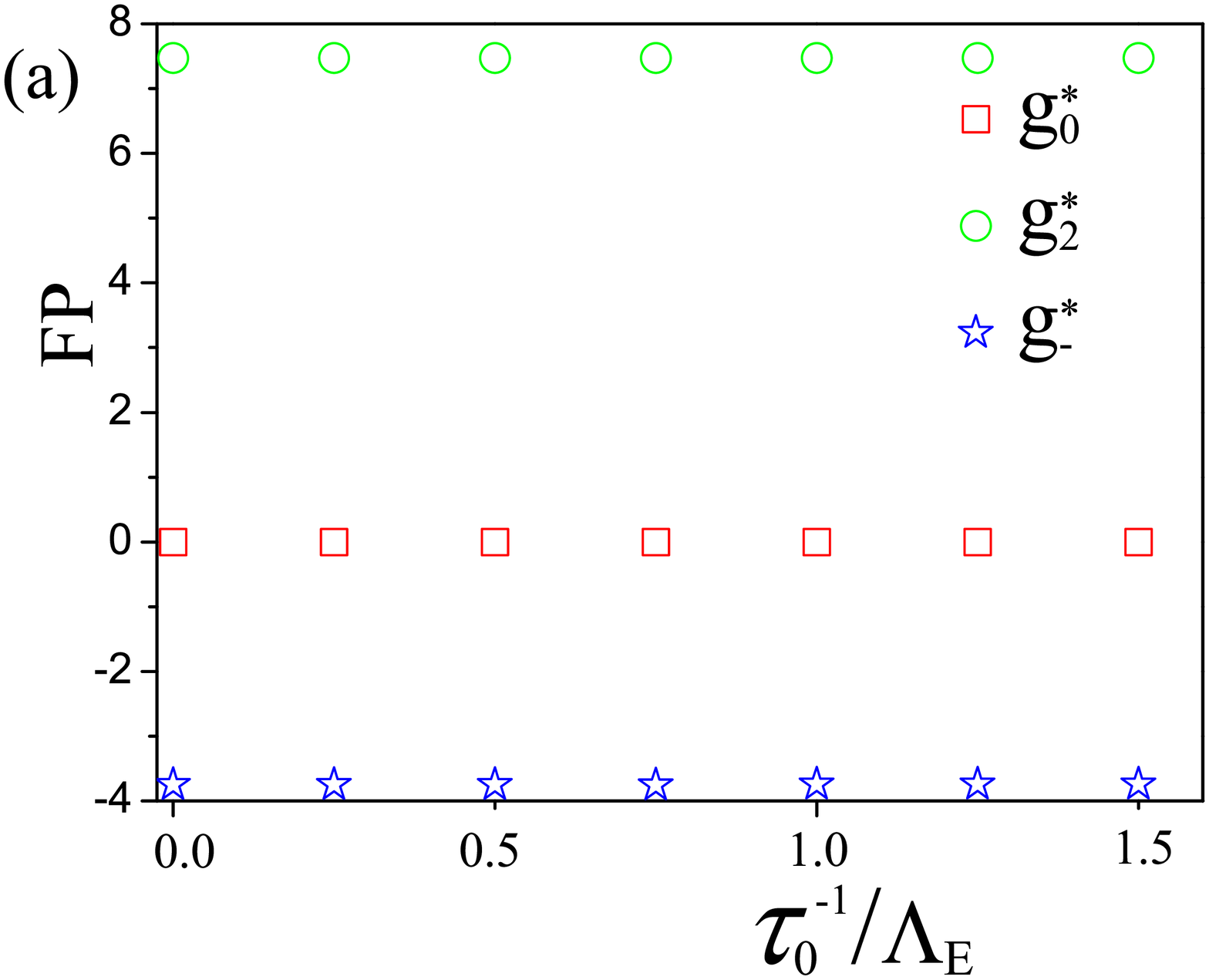}\hspace{-0.80cm}
\includegraphics[width=3.6in]{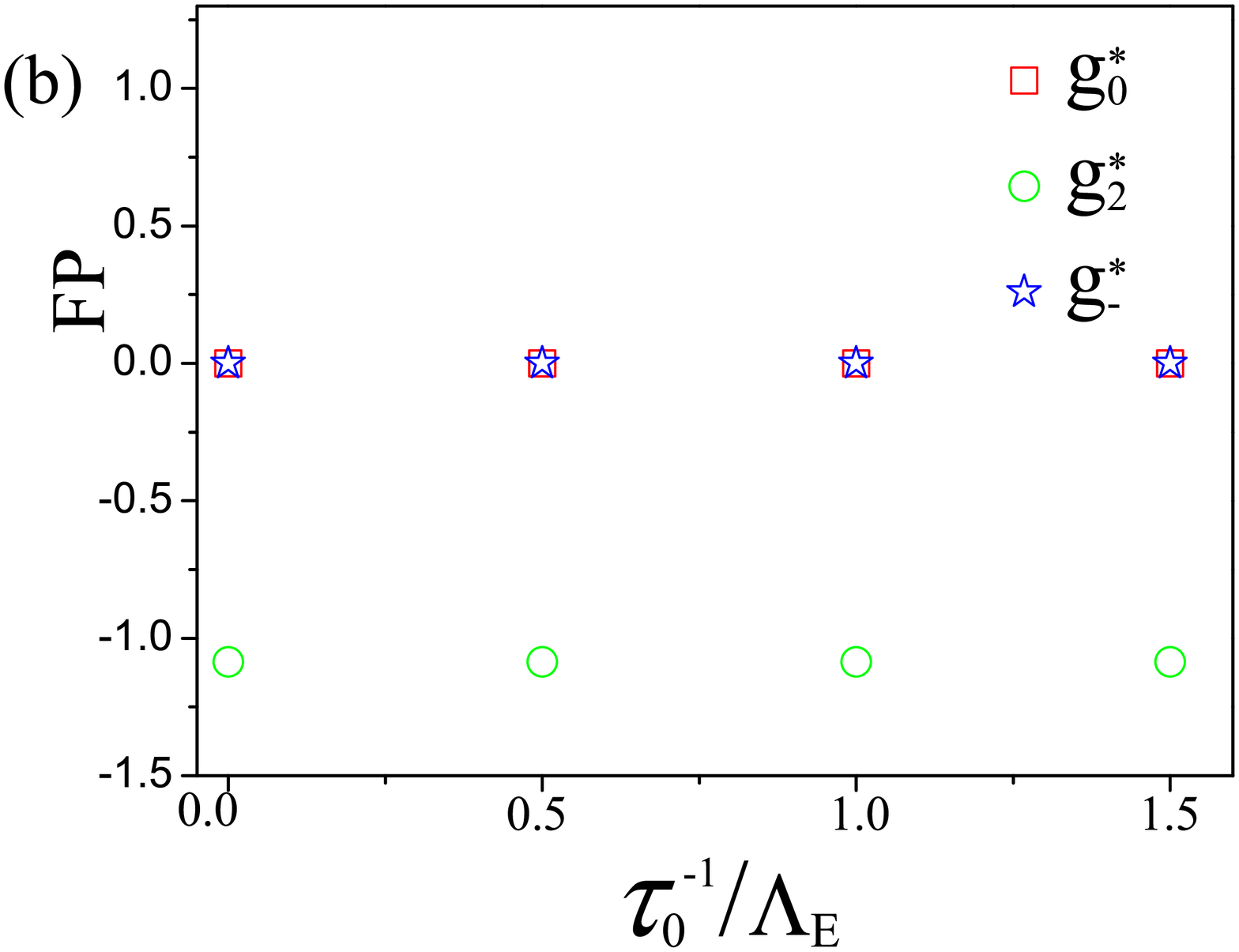}
\vspace{-0.36cm}
\caption{(Color online) \textcolor{black}{Evolution of the fixed points $(g^*_0,g^*_-,g^*_2)$ for the checkerboard lattice
in the presence of random gauge potential around (a): the QAH fixed point $(0,-3.73,7.46)g_+$
and (b): the QSH fixed point $(0,0,-1.09)g_+$ at some representatively initial values of
disorder strength of the random chemical potential. FP represents the fixed points
and the parameter The parameter $\tau^{-1}\sim n_0\nu^2_m/t$ designates the impurity scattering rate and
$\Lambda_E=t\Lambda^2$ with $t$ rescaled by $\hbar^2/2m$.}}\label{Fig_M13_dis}
\end{figure}

\begin{figure}
\centering
\includegraphics[width=3.0in]{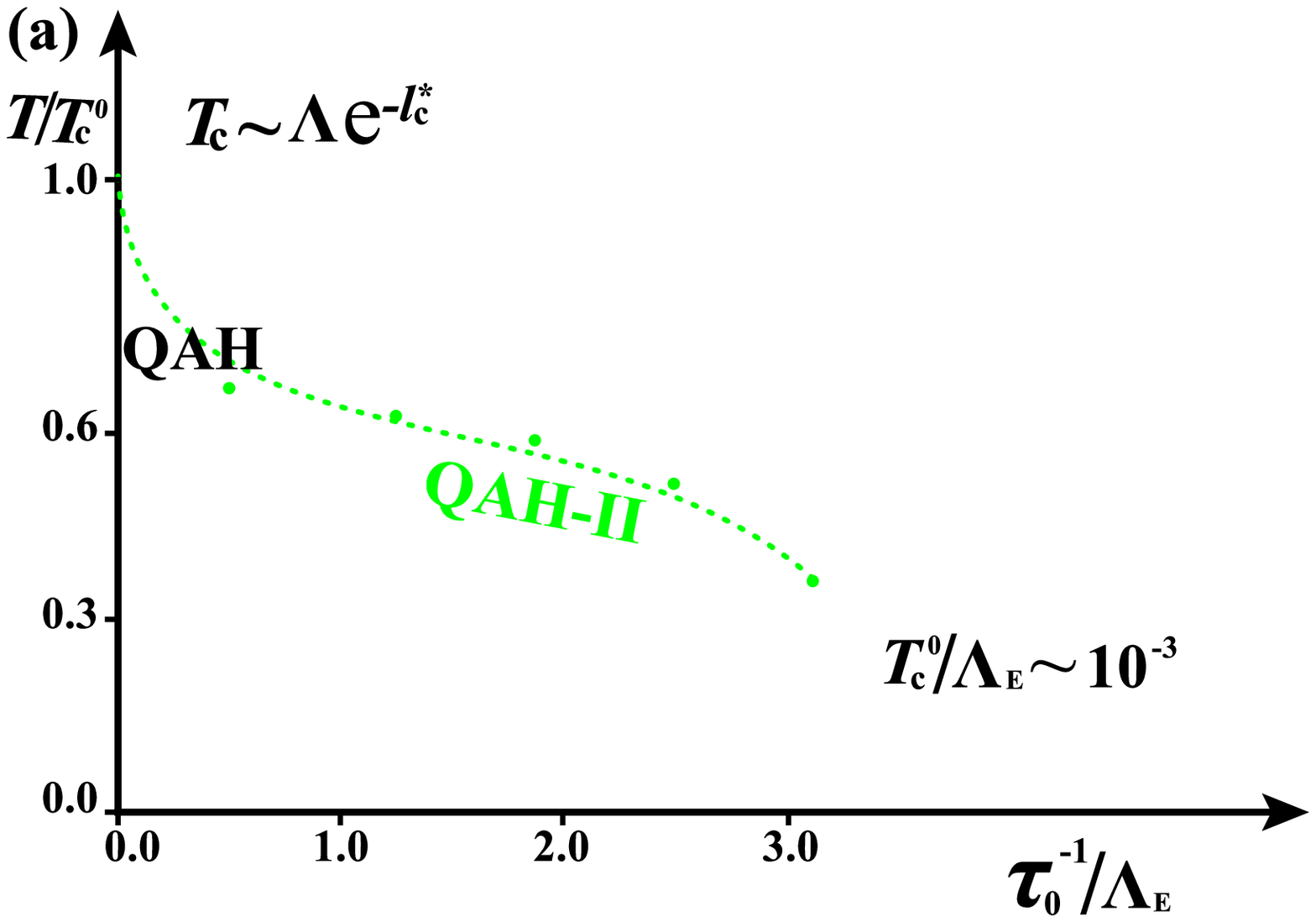}\hspace{0.80cm}
\includegraphics[width=3.0in]{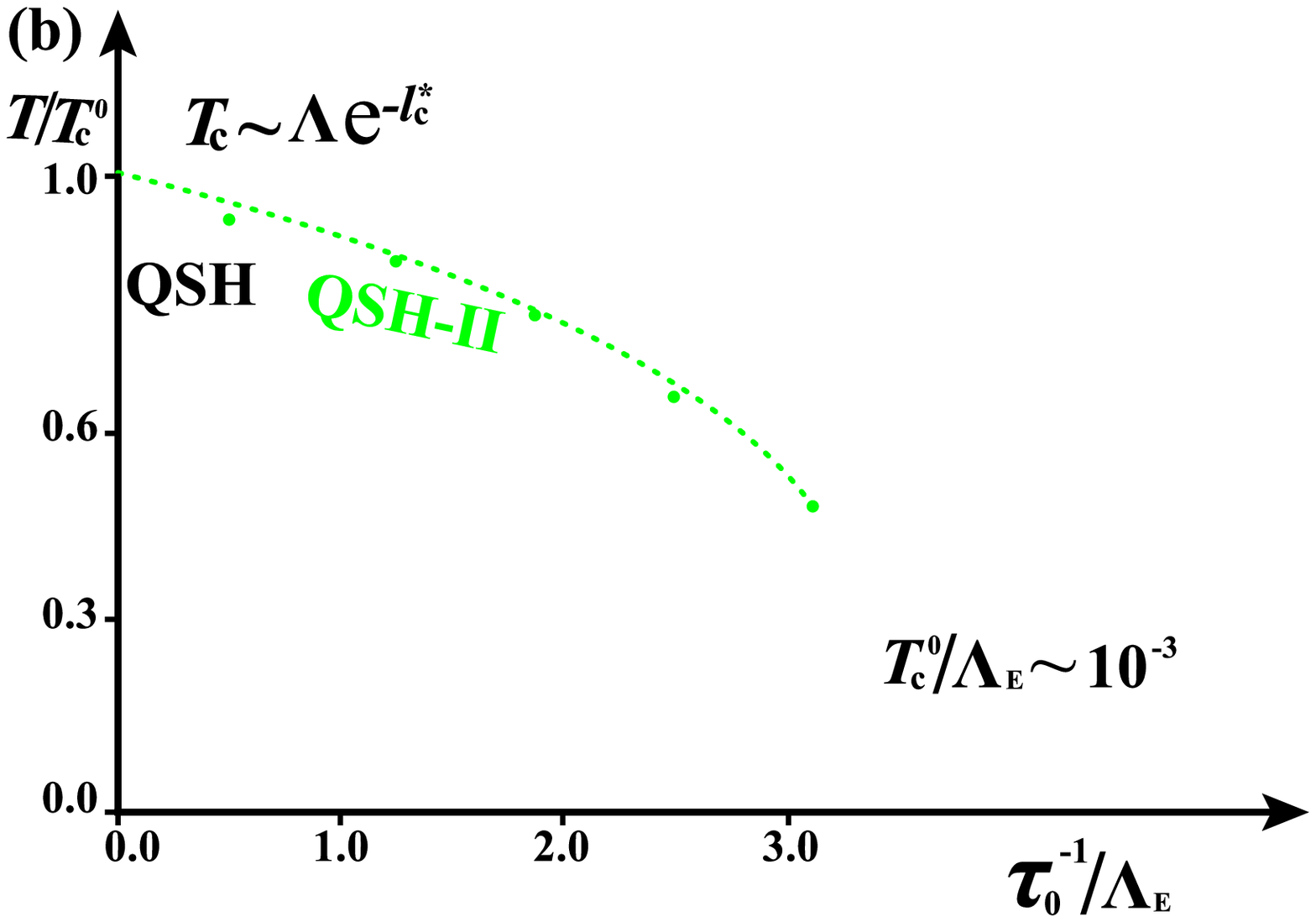}
\vspace{-0.2cm}
\caption{(Color online) \textcolor{black}{Schematic phase diagram for the checkerboard lattice in the presence
of random mass potential and which in the clean limit give  (a): $(g^*_0,g^*_-,g^*_2)=(0,-3.73,7.46)g^+$ and
(b): $(g^*_0,g^*_-,g^*_2)=(0,0,-1.09)g^+$. The parameter $\tau^{-1}\sim n_0\nu^2_m/t$ designates the impurity scattering rate and
$\Lambda_E=t\Lambda^2$ with $t$ rescaled by $\hbar^2/2m$. The phases QAH (or QSH) and QAH-II (QSH-II) are the same phase
but with different critical temperatures caused by distinct of FPs as depicted in Fig.~\ref{Fig_M2_dis}.}}\label{Fig_Phase_diagram_mass}
\end{figure}

\begin{figure}
\centering
\includegraphics[width=3.0in]{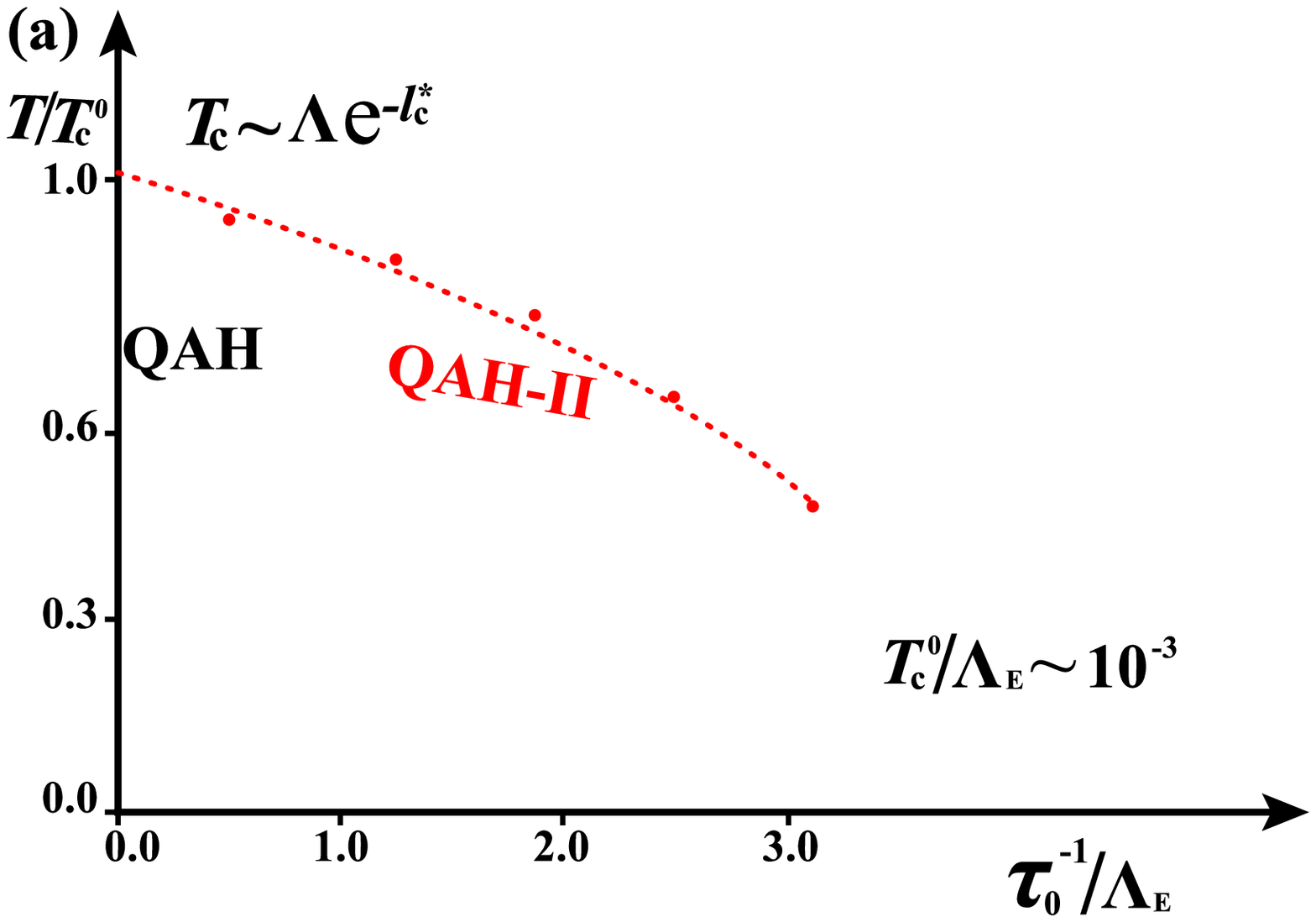}\hspace{0.80cm}
\includegraphics[width=3.0in]{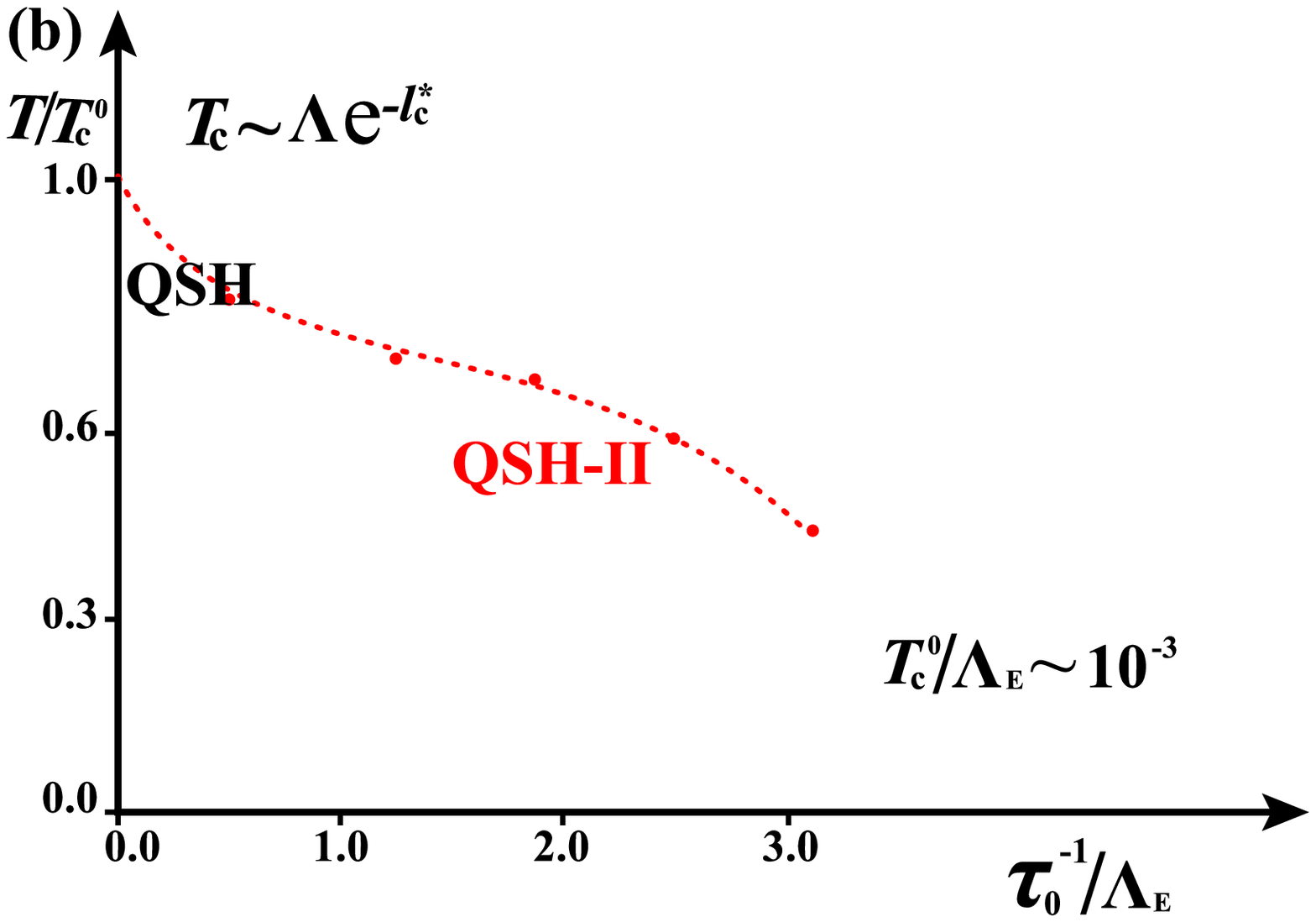}
\vspace{-0.2cm}
\caption{(Color online) \textcolor{black}{Schematic phase diagram for the checkerboard lattice in the presence
of random gauge potential and which in the clean limit give (a): $(g^*_0,g^*_-,g^*_2)=(0,-3.73,7.46)g^+$ and
(b): $(g^*_0,g^*_-,g^*_2)=(0,0,-1.09)g^+$. The parameter $\tau^{-1}\sim n_0\nu^2_m/t$ designates the impurity scattering rate.
The phases QAH (or QSH) and QAH-II (QSH-II) are the same phase but with different critical temperatures
caused by distinct of FPs as depicted in Fig.~\ref{Fig_M13_dis}.}}\label{Fig_Phase_diagram_gauge}
\end{figure}

\begin{figure}
\centering
\includegraphics[width=3.8in]{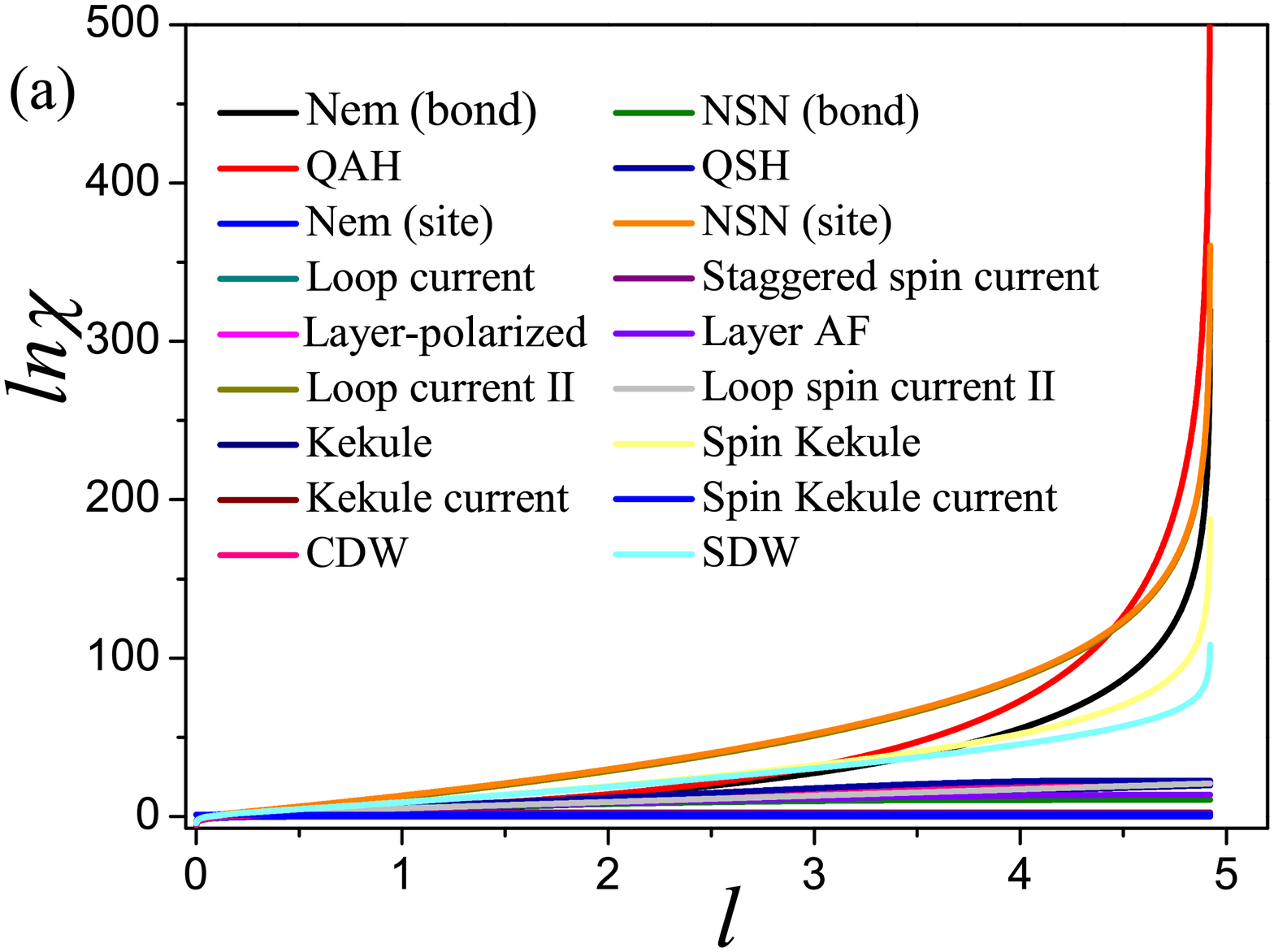}\hspace{-1.6cm}
\includegraphics[width=3.8in]{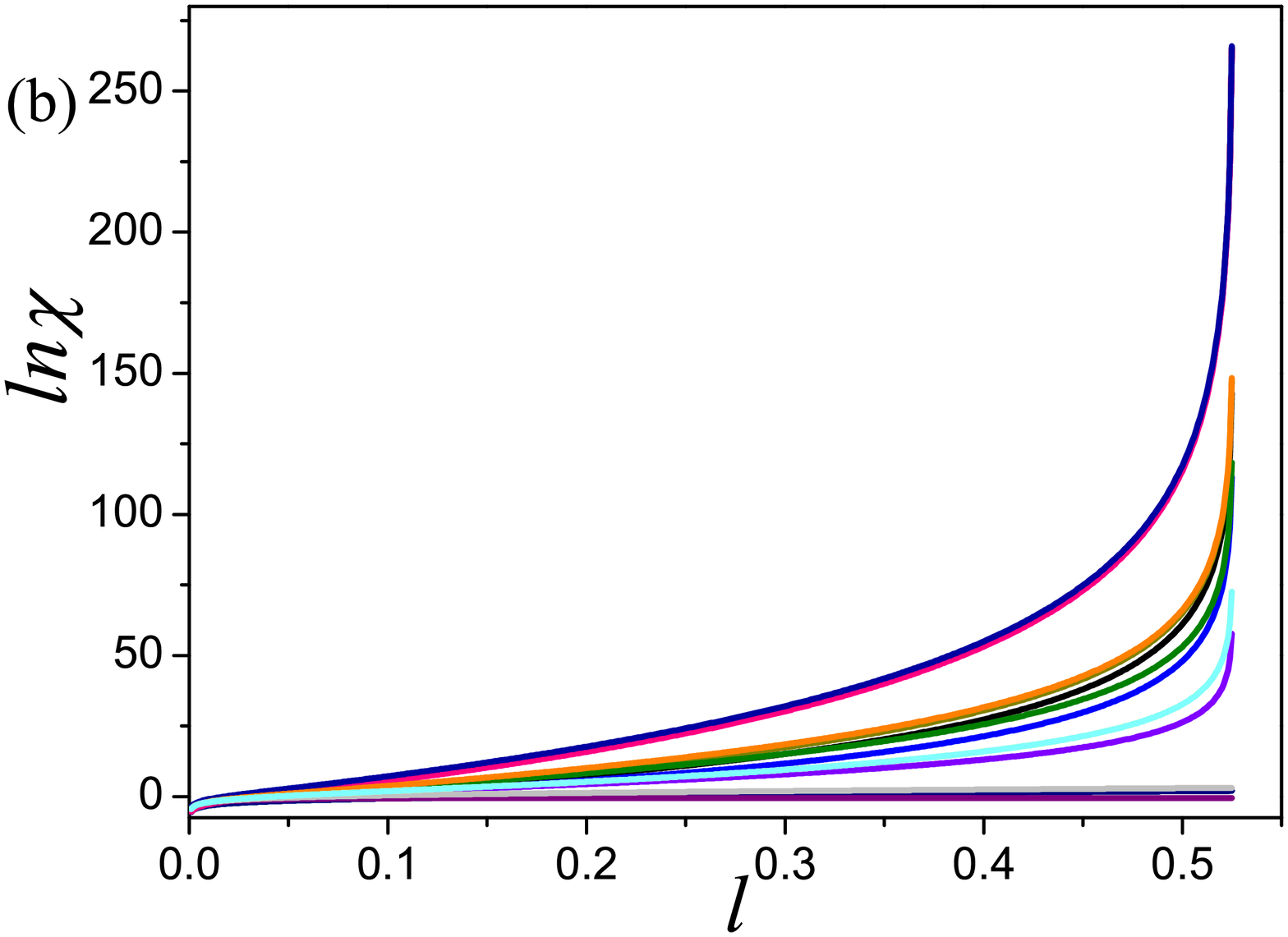}
\includegraphics[width=3.8in]{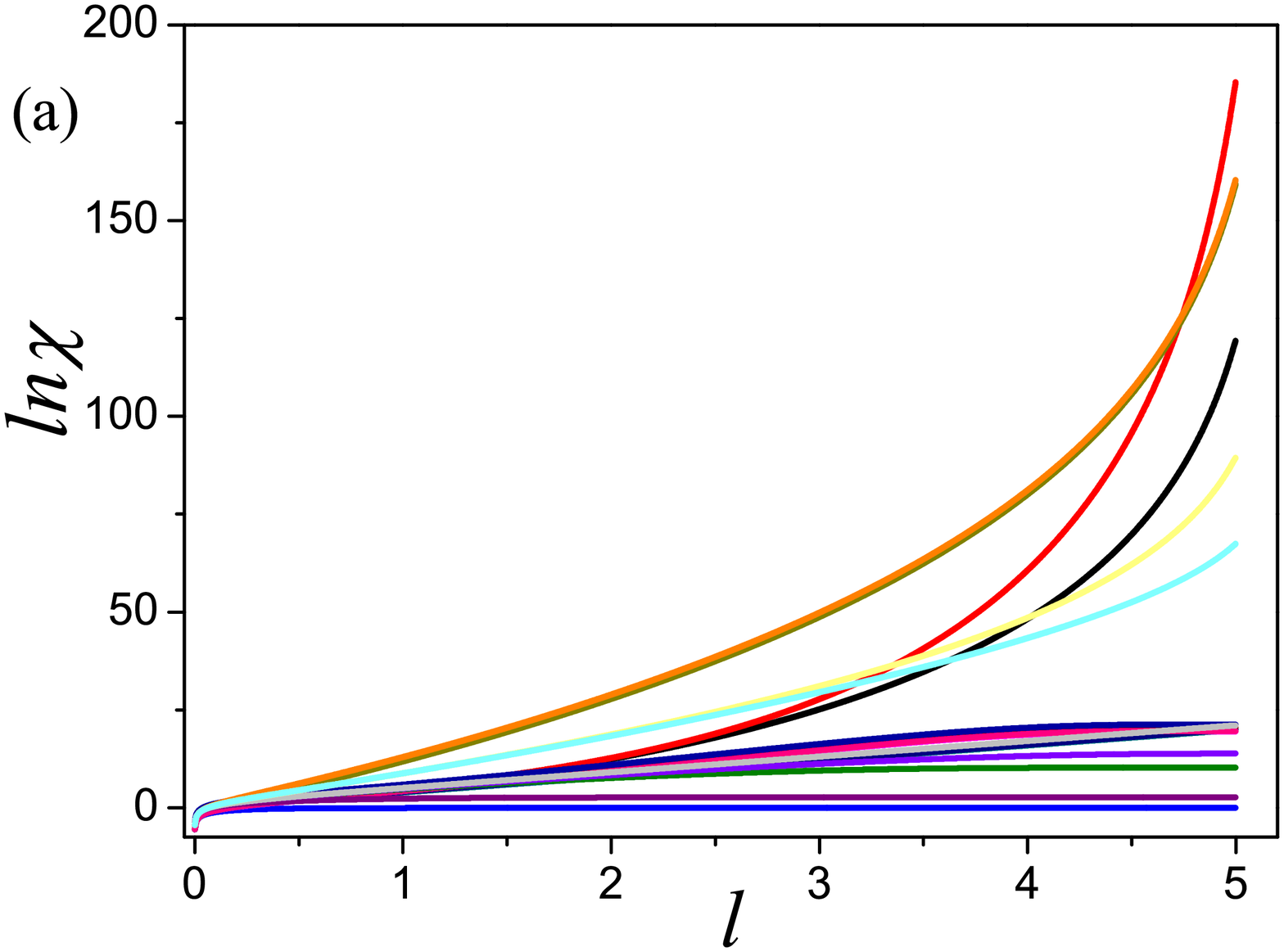}\hspace{-1.6cm}
\includegraphics[width=3.8in]{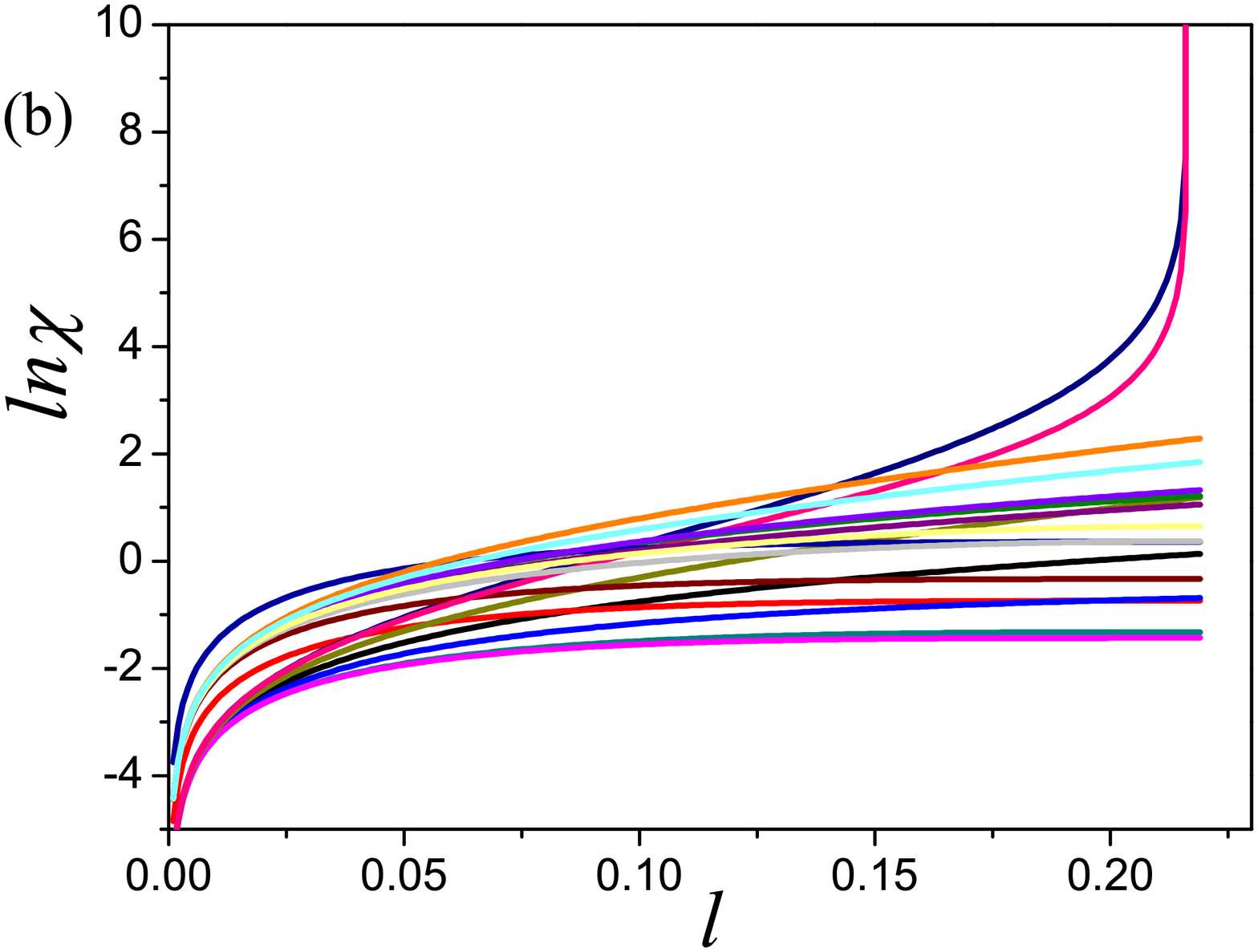}
\vspace{-0.36cm}
\caption{(Color online) Flows of all charge and spin susceptibilities for the bilayer honeycomb lattice. UP: in clean limit
in the vicinity of the fixed point QAH $(g^*_0,g^*_-,g^*_2)=(0,-3.73,7.46)g^+$ (a) and QSH  $(g^*_0,g^*_-,g^*_2)=(0,0,-1.09)g^+$ (b);
DOWN: in the presence of random chemical potential in the vicinity of the QAH fixed point $(g^*_0,g^*_-,g^*_2)=(0,-3.73,7.46)g^+$,
(a): small bare $v^0_m$. The leading instability is QAH; (b): large bare $v^0_m$. The leading instability is CDW.
Susceptibilities for other fixed points and disorders can be derived similarly and are
not shown here.}\label{Fig_chi_Honeycomb}
\end{figure}

After fulfilling the one-loop corrections to the source terms~\ref{Fig_fermion_source_correction},
we can get the flow equations of the source terns \cite{Vafek2014PRB}
(the matrixes $M^s_0=1\vec{s}$, $M^s_1=\tau_1\vec{s}$, $M^s_2=\tau_2\vec{s}$,
and $M^s_3=\tau_3\vec{s}$ refer to the FM, NSN (bond), QSH, and NSN (site),
respectively)£º
i) random chemical potential: $M=\tau_0$
\begin{eqnarray}
\frac{d\ln\Delta^{c}_1}{dl}&=&2+\frac{4m}{16\pi}g_0-\frac{12m}{16\pi}g_1-\frac{4m}{16\pi}g_2-\frac{4m}{16\pi}g_3,\\
\frac{d\ln\Delta^{c}_2}{dl}&=&2+\frac{8m}{16\pi}g_0-\frac{8m}{16\pi}g_1-\frac{24m}{16\pi}g_2-\frac{8m}{16\pi}g_3
-\frac{n_0v^2_m}{2\pi t^2},\\
\frac{d\ln\Delta^{c}_3}{dl}&=&2+\frac{4m}{16\pi}g_0-\frac{4m}{16\pi}g_1-\frac{4m}{16\pi}g_2-\frac{12m}{16\pi}g_3,
\end{eqnarray}
and
\begin{eqnarray}
\frac{d\ln\Delta^{s}_1}{dl}&=&2+\frac{4m}{16\pi}g_0+\frac{4m}{16\pi}g_1-\frac{4m}{16\pi}g_2-\frac{4m}{16\pi}g_3,\\
\frac{d\ln\Delta^{s}_2}{dl}&=&2+\frac{8m}{16\pi}g_0-\frac{8m}{16\pi}g_1+\frac{8m}{16\pi}g_2-\frac{8m}{16\pi}g_3
-\frac{n_0v^2_m}{2\pi t^2},\\
\frac{d\ln\Delta^{s}_3}{dl}&=&2+\frac{4m}{16\pi}g_0-\frac{4m}{16\pi}g_1-\frac{4m}{16\pi}g_2+\frac{4m}{16\pi}g_3;
\end{eqnarray}
ii) Random gauge potential: $M=\tau_1$ and $M=\tau_3$
\begin{eqnarray}
\frac{d\ln\Delta^{c}_1}{dl}&=&2+\frac{4m}{16\pi}g_0-\frac{12m}{16\pi}g_1-\frac{4m}{16\pi}g_2-\frac{4m}{16\pi}g_3,\\
\frac{d\ln\Delta^{c}_2}{dl}&=&2+\frac{8m}{16\pi}g_0-\frac{8m}{16\pi}g_1-\frac{24m}{16\pi}g_2-\frac{8m}{16\pi}g_3
+\frac{n_0v^2_m}{2\pi t^2},\\
\frac{d\ln\Delta^{c}_3}{dl}&=&2+\frac{4m}{16\pi}g_0-\frac{4m}{16\pi}g_1-\frac{4m}{16\pi}g_2-\frac{12m}{16\pi}g_3,
\end{eqnarray}
and
\begin{eqnarray}
\frac{d\ln\Delta^{s}_1}{dl}&=&2+\frac{4m}{16\pi}g_0+\frac{4m}{16\pi}g_1-\frac{4m}{16\pi}g_2-\frac{4m}{16\pi}g_3,\\
\frac{d\ln\Delta^{s}_2}{dl}&=&2+\frac{8m}{16\pi}g_0-\frac{8m}{16\pi}g_1+\frac{8m}{16\pi}g_2-\frac{8m}{16\pi}g_3
+\frac{n_0v^2_m}{2\pi t^2},\\
\frac{d\ln\Delta^{s}_3}{dl}&=&2+\frac{4m}{16\pi}g_0-\frac{4m}{16\pi}g_1-\frac{4m}{16\pi}g_2+\frac{4m}{16\pi}g_3;
\end{eqnarray}
iii) Random mass: $M=\tau_2$
\begin{eqnarray}
\frac{d\ln\Delta^{c}_1}{dl}&=&2+\frac{4m}{16\pi}g_0-\frac{12m}{16\pi}g_1-\frac{4m}{16\pi}g_2-\frac{4m}{16\pi}g_3,\\
\frac{d\ln\Delta^{c}_2}{dl}&=&2+\frac{8m}{16\pi}g_0-\frac{8m}{16\pi}g_1-\frac{24m}{16\pi}g_2-\frac{8m}{16\pi}g_3
-\frac{n_0v^2_m}{2\pi t^2},\\
\frac{d\ln\Delta^{c}_3}{dl}&=&2+\frac{4m}{16\pi}g_0-\frac{4m}{16\pi}g_1-\frac{4m}{16\pi}g_2-\frac{12m}{16\pi}g_3,
\end{eqnarray}
and
\begin{eqnarray}
\frac{d\ln\Delta^{s}_1}{dl}&=&2+\frac{4m}{16\pi}g_0+\frac{4m}{16\pi}g_1-\frac{4m}{16\pi}g_2-\frac{4m}{16\pi}g_3,\\
\frac{d\ln\Delta^{s}_2}{dl}&=&2+\frac{8m}{16\pi}g_0-\frac{8m}{16\pi}g_1+\frac{8m}{16\pi}g_2-\frac{8m}{16\pi}g_3
-\frac{n_0v^2_m}{2\pi t^2},\\
\frac{d\ln\Delta^{s}_3}{dl}&=&2+\frac{4m}{16\pi}g_0-\frac{4m}{16\pi}g_1-\frac{4m}{16\pi}g_2+\frac{4m}{16\pi}g_3.
\end{eqnarray}

\textcolor{black}{Some of the corresponding results are provided in Figs.~\ref{Fig_M0_dis}, \ref{Fig_M2_dis}, \ref{Fig_M13_dis},
\ref{Fig_Phase_diagram_mass}, and \ref{Fig_Phase_diagram_gauge} besides the figures presented in the main text.}

\section{Results for the bilayer honeycomb lattice}

\subsection{Effective theory for bilayer graphene}

The tight-binding Hamiltonian for electrons hopping on the bilayer honeycomb lattice with Bernal stacking can be
described as \cite{Vafek2010PRB,Vafek2010PRB_2},
\begin{eqnarray}
H=\sum_{\mathbf{r}\mathbf{r'}}\left[t_{\mathbf{r}\mathbf{r'}}c^\dagger_\sigma(\mathbf{r})c_\sigma(\mathbf{r'})+
\mathrm{H.c.}\right]+\frac{1}{2}\sum_{\mathbf{r}\mathbf{r'}}\delta n{\mathbf{r}}V(\mathbf{r}-\mathbf{r'})
\delta n(\mathbf{r'}),
\end{eqnarray}
where $t$ represents the hopping amplitudes connecting the next-nearest sites in a plan and $n(\mathbf{r})=c^\dagger(\mathbf{r})
c(\mathbf{r})$. We then transfer above Hamiltonian to momentum space
and gain \cite{McCann2006PRL,Nilsson2008PRB,Vafek2010PRB,Vafek2010PRB_2},
\begin{eqnarray}
H=\sum_{\mathbf{k}}\psi^\dagger_{\mathbf{k}\sigma}\mathcal{H}_0\psi_{\mathbf{k}\sigma},
\end{eqnarray}
where $\psi^\dagger=(a^\dagger_{1,\mathbf{k}},a^\dagger_{2,\mathbf{k}},b^\dagger_{2,\mathbf{k}},b^\dagger_{1,\mathbf{k}})$
and $\mathcal{H}_0$ can be described as \cite{McCann2006PRL,Nilsson2008PRB,Vafek2010PRB,Vafek2010PRB_2}
\begin{eqnarray}
\mathcal{H}_0=\left(
  \begin{array}{cccc}
    0 & d^*_{\mathbf{k}} & t_\perp & 0 \\
    d_{\mathbf{k}} & 0 & 0 & 0 \\
    t_\perp & 0 & 0 & d_{\mathbf{k}} \\
    0 & 0 & d^*_{\mathbf{k}} & 0 \\
  \end{array}
\right)
\end{eqnarray}
which can straightforwardly be diagonalized to \cite{Vafek2010PRB,Vafek2010PRB_2}
\begin{eqnarray}
E(\mathbf{k})=\pm\left(\frac{1}{2}t_\perp\pm\sqrt{|d_{\mathbf{k}}|^2+\frac{1}{4}t^2_\perp}\right)
\end{eqnarray}
with
\begin{eqnarray}
d_{\mathbf{k}}=t\left[2\cos\left(\frac{\sqrt{3}}{2}k_ya\right)e^{-\frac{i}{2}k_xa}+e^{ik_xa}\right].
\end{eqnarray}
Two of these four bands are parabolically touching at $\mathbf{k}=0$ \cite{Vafek2010PRB}, which can also
considered as a QBCP system and whose band structure is similar to the checkerboard's.
The effective action of noninteracting terms in clean limit for the bilayer graphene
would be given by \cite{McCann2006PRL,Nilsson2008PRB,Vafek2010PRB}
\begin{eqnarray}
S_0=\int d\tau\left\{\sum_{|\mathbf{k}|<\Lambda}\sum_{\sigma=\uparrow\downarrow}\psi^\dagger_{\mathbf{k}\sigma}
\left[\partial_\tau+\frac{k^2_x-k^2_y}{2m}\tau_0\sigma_1+\frac{2k_xk_y}{2m}\tau_3\sigma_2\right]\psi_{\mathbf{k}\sigma}\right\}.
\end{eqnarray}
The Pauli matrices $\sigma^i$ act on the layer indices 1-2 and the $\tau$ matrices act on the valley indices $\mathbf{K}-\mathbf{K'}$.
The effective mass is $m=\frac{2t_\perp}{9t^2}$ and $\psi$ represents $\frac{N}{2}$ copies of the four component pseudospinor.
$N=4$ for spin 1/2. All marginal interactions are
\begin{eqnarray}
S_{\mathrm{int}}
&=&\int d\tau\int d^2\mathbf{x}\left\{\frac{2\pi}{m}g_0\left(\sum_{\sigma=\uparrow\downarrow}\psi^\dagger_\sigma
(\mathbf{x})\tau_0\sigma_0\psi_\sigma(\mathbf{x})\right)^2+\frac{2\pi}{m}g_1\left(\sum_{\sigma=\uparrow\downarrow}\psi^\dagger_\sigma
(\mathbf{x})\tau_0\sigma_1\psi_\sigma(\mathbf{x})\right)^2\right.\nonumber\\
&&\left.+\frac{2\pi}{m}g_2\left(\sum_{\sigma=\uparrow\downarrow}\psi^\dagger_\sigma
(\mathbf{x})\tau_3\sigma_2\psi_\sigma(\mathbf{x})\right)^2
+\frac{2\pi}{m}g_3\left(\sum_{\sigma=\uparrow\downarrow}\psi^\dagger_\sigma
(\mathbf{x})\tau_3\sigma_3\psi_\sigma(\mathbf{x})\right)^2\right\}.
\end{eqnarray}

In order to compare with the effective theory of the checkerboard lattice, we introduce
the new "Pauli" matrices by defining $\sigma'_0\equiv\tau_0\sigma_0$, $\sigma'_1\equiv\tau_3\sigma_2$,
$\sigma'_2\equiv\tau_3\sigma_3$ and $\sigma'_3\equiv\tau_0\sigma_1$, which also have the same symmetries
of the Pauli matrices as $\sigma'_\mu\sigma'_\nu=1_4\delta_{\mu\nu}+i\epsilon_{\mu\nu\lambda}\sigma'_\lambda$.
We finally obtain the effective action in the presence of disorders after carrying out the fourier
transformation,
\begin{eqnarray}
S_{\mathrm{eff}}&=&\int^{+\infty}_{-\infty}\frac{d\omega}{2\pi}\int^{\Lambda}\frac{d^2\mathbf{k}}{(2\pi)^2}\sum_{\sigma=\uparrow\downarrow}
\psi^\dagger_{\sigma}(\omega,\mathbf{k})[-i\omega+2tk_xk_y\sigma'_1+t(k^2_x-k^2_y)\sigma'_3]\psi_{\sigma}(\omega,\mathbf{k})
+4\pi t\sum^{3}_{i=0}g_i\int^{+\infty}_{-\infty}\frac{d\omega_1d\omega_2d\omega_3}{(2\pi)^3}\nonumber\\
&&\times\int^{\Lambda}\frac{d^2\mathbf{k}_1d^2\mathbf{k}_2d^2\mathbf{k}_3}{(2\pi)^6}\sum_{\sigma,\sigma'=\uparrow\downarrow}
\psi^\dagger_\sigma(\omega_1,\mathbf{k}_1)\sigma'_i\psi_\sigma(\omega_2,\mathbf{k}_2)\psi^\dagger_{\sigma'}(\omega_3,\mathbf{k}_3)
\sigma'_i\psi_{\sigma'}(\omega_1+\omega_2-\omega_3,\mathbf{k}_1+\mathbf{k}_2-\mathbf{k}_3)\nonumber\\
&&+n_0\nu_m\int^{+\infty}_{-\infty}\frac{d\omega}{2\pi}\int \frac{d^2\mathbf{k}d^2\mathbf{k'}}{(2\pi)^4}\psi^\dagger(\mathbf{k},\omega)
M\psi(\mathbf{k'},\omega)A(\mathbf{k-k'}),\label{Eq_effective_action}
\end{eqnarray}
with $M=\sigma'_0$ being the random chemical potential, $M=\sigma'_1$ and $M=\sigma'_3$ being the random gauge potential
(two components), and $M=\sigma'_2$ being the random mass.

\subsection{Fixed points and susceptibilities for the bilayer honeycomb lattice in the presence of disorder}

We emphasize that there are 12 other possible orders besides the 6 orders in the checkerboard lattice for the
honeycomb lattice as provided in Ref. \cite{Vafek2012PRB}, which are listed here for completeness:
\vskip 0.5cm
\centerline{
\begin{tabular}{|l|l|}
                \hline
   Charge channel & Spin channel \\
   \hline
   $\tau_0\otimes\sigma_0$: charge instability & $\tau_0\otimes\sigma_0\vec{s}$: FM \\
   $\tau_0\otimes\sigma_1$: nematic (bond) \cite{Vafek2010PRB,Lemonik2010PRB} ~~~~~~~~~~~~
  & $\tau_0\otimes\sigma_1\vec{s}$: NSN (bond) \\
  $\tau_3\otimes\sigma_3$: QAH \cite{Haldane1988PRL,Nandkishore2010PRB}
  & $\tau_3\otimes\sigma_3\vec{s}$: QSH \cite{Throckmorton2012PRB,Lemonik2012PRB,Scherer2012PRB} \\
  $\tau_3\otimes\sigma_2$: nematic (site) & $\tau_3\otimes\sigma_2\vec{s}$: NSN (site) \\
   $\tau_3\otimes\sigma_0$: Loop current \cite{Varma2013PRB} &
  $\tau_3\otimes\sigma_0\vec{s}$: Staggered spin current ~~~~~~~~~~~~\\
   $\tau_0\otimes\sigma_3$: Layer-polarized \cite{MacDonald2008PRB,Nandkishore2010PRL} &
  $\tau_0\otimes\sigma_3\vec{s}$: Layer AF \cite{Neto2009RMP,Vafek2010PRB_2,Kharitonov2012PRB} \\
   $\tau_3\otimes\sigma_1$: Loop current II~~~~~~~~~~ 
 & $\tau_3\otimes\sigma_1\vec{s}$: Loop spin current II  \\ 
   $\tau_1\otimes\sigma_1$: Kekul\'{e} \cite{Mudry2007PRL} & $\tau_1\otimes\sigma_1\vec{s}$: Spin Kekul\'{e} \\
   $\tau_1\otimes\sigma_2$: Kekul\'{e} current & $\tau_1\otimes\sigma_2\vec{s}$: Spin Kekul\'{e} current\\
   $\tau_1\otimes\sigma_0$: CDW & $\tau_1\otimes\sigma_0\vec{s}$: SDW \\
      \hline
               \end{tabular}
}
\vskip 0.5cm
Additional, there is the third fixed point in the honeycomb lattice besides the two fixed points considered
in the checkerboard lattice due to the nonzero initial values of $g_1$ and $g_2$. For instance, the fixed point $(g^*_0,g^*_-,g^*_2)=(0,3.73,7.46)g^+$~\cite{Vafek2014PRB}. By considering all these facets and paralleling the
similar steps employed in the checkerboard lattice, we provide the primary results of the susceptibilities for both clean limit and
impurity case in the vicinity of the some representative fixed points as presented in Fig.~\ref{Fig_chi_Honeycomb}.
We summarize all the information from the susceptibilities and plot the schematic phase diagram, namely Fig. 5 in the main text.

\end{document}